\documentclass[10pt,conference,letterpaper]{IEEEtran}

\pdfoutput=1 

\newcommand{\seq}[1]{{$\langle\,${\small \textsf{\textbf{#1}}}$\,\rangle$}}
\newcommand{\term}[1]{{\small \textsf{\textbf{#1}}}}
\newcommand{\smalltt}[1]{{\texttt{\small {#1}}}}

\renewcommand{\baselinestretch}{0.99}

\usepackage{times}
\usepackage{amsmath}
\usepackage[ruled,vlined,linesnumberedhidden]{algorithm2e}
\usepackage{subfigure}
\usepackage{balance}
\usepackage{booktabs}
\usepackage{datetime}
\usepackage{graphicx}

\title{Computing n-Gram Statistics in MapReduce}

\author{
{Klaus Berberich{\small $~^{\#}$}, Srikanta Bedathur{\small $~^{*}$}}
\vspace{1mm}\\
\fontsize{10}{10}\selectfont\itshape
$~^{\#}$Max Planck Institute for Informatics\\
Saarbr\"ucken, Germany\\
\fontsize{9}{9}\selectfont\ttfamily\upshape
$~^{1}$kberberi@mpi-inf.mpg.de
\vspace{1mm}\\
\fontsize{10}{10}\selectfont\rmfamily\itshape
$~^{*}$Indraprastha Institute of Information Technology\\
New Delhi, India\\
\fontsize{9}{9}\selectfont\ttfamily\upshape
$~^{2}$bedathur@iiitd.ac.in
}
\begin{document}
\maketitle
\begin{abstract} 
  Statistics about $n$-grams (i.e., sequences of contiguous words or
  other tokens in text documents or other string data) are an
  important building block in information retrieval and natural
  language processing. In this work, we study how $n$-gram statistics,
  optionally restricted by a maximum $n$-gram length and minimum
  collection frequency, can be computed efficiently harnessing
  MapReduce for distributed data processing. We describe different
  algorithms, ranging from an extension of word counting, via methods
  based on the \textsc{Apriori} principle, to a novel method
  \textsc{Suffix}-$\sigma$ that relies on sorting and aggregating
  suffixes. We examine possible extensions of our method to support
  the notions of maximality/closedness and to perform aggregations
  beyond occurrence counting. Assuming Hadoop as a concrete MapReduce
  implementation, we provide insights on an efficient implementation
  of the methods. Extensive experiments on The New York Times
  Annotated Corpus and ClueWeb09 expose the relative benefits and
  trade-offs of the methods.
\end{abstract}

\section{Introduction}
\label{sec:introduction}

Applications in various fields including information
retrieval~\cite{Bernstein:2006uq,Zhai:2008fk} and natural language
processing~\cite{Brants:2007ys,Federico:2008fk,Stolcke:2002kx} rely on
statistics about $n$-grams (i.e., sequences of contiguous words in
text documents or other string data) as an important building
block. Google and Microsoft have made available $n$-gram statistics
computed on parts of the Web. While certainly a valuable resource, one
limitation of these datasets is that they only consider $n$-grams
consisting of up to five words. With this limitation, there is no way
to capture idioms, quotations, poetry, lyrics, and other types of
named entities (e.g., products, books, songs, or movies) that
typically consist of more than five words and are crucial to
applications including plagiarism detection, opinion mining, and
social media analytics.

MapReduce has gained popularity in recent years both as a programming
model and in its open-source implementation Hadoop. It provides a
platform for distributed data processing, for instance, on web-scale
document collections. MapReduce imposes a rigid programming model, but
treats its users with features such as handling of node failures and
an automatic distribution of the computation. To make most effective
use of it, problems need to be cast into its programming model, taking
into account its particularities.

In this work, we address the problem of efficiently computing $n$-gram
statistics on MapReduce platforms. We allow for a restriction of the
$n$-gram statistics to be computed by a maximum length $\sigma$ and a
minimum collection frequency $\tau$. Only $n$-grams consisting of up
to $\sigma$ words and occurring at least $\tau$ times in the document
collection are thus considered.

While this can be seen as a special case of frequent sequence mining,
our experiments on two real-world datasets show that MapReduce
adaptations of \textsc{Apriori}-based
methods~\cite{Srikant:1996fk,Zaki:2001ly} do not perform well -- in
particular when long and/or less frequent $n$-grams are of
interest. In this light, we develop our novel method
\textsc{Suffix}-$\sigma$ that is based on ideas from string
processing. Our method makes thoughtful use of MapReduce's grouping
and sorting functionality. It keeps the number of records that have to
be sorted by MapReduce low and exploits their order to achieve a
compact main-memory footprint, when determining collection frequencies
of all $n$-grams considered.

We also describe possible extensions of our method. This includes the
notions of maximality/closedness, known from frequent sequence mining,
that can drastically reduce the amount of $n$-gram statistics
computed. In addition, we investigate to what extent our method can
support aggregations beyond occurrence counting, using $n$-gram time
series, recently made popular by Michel et al.~\cite{Michel:ys}, as an
example.

\textbf{Contributions} made in this work include:
\begin{itemize}
\vspace{-1mm}
\item a \textit{novel method} \textsc{Suffix}-$\sigma$ to compute
  $n$-gram statistics that has been specifically designed for
  MapReduce;

\item a detailed account on \textit{efficient implementation} and
  \textit{possible extensions} of \textsc{Suffix}-$\sigma$ (e.g., to
  consider maximal/closed $n$-grams or support other aggregations);

\item a \textit{comprehensive experimental evaluation} on The New York
  Times Annotated Corpus (1.8 million news articles from 1987--2007)
  and ClueWeb09-B (50 million web pages crawled in 2009), as two
  large-scale real-world datasets, comparing our method against
  state-of-the-art competitors and investigating their trade-offs.
\vspace{-1mm}
\end{itemize}

\textsc{Suffix}-$\sigma$ outperforms its best competitor in our
experiments by up to a \textit{factor 12x} when long and/or less
frequent $n$-grams are of interest. Otherwise, it performs at least on
par with the best competitor.

\textbf{Organization.} Section~\ref{sec:model--background} introduces
our model. Section~\ref{sec:methods-based-prior} details on methods to
compute $n$-gram statistics based on prior
ideas. Section~\ref{sec:suffix-sorting} introduces our method
\textsc{Suffix}-$\sigma$. Aspects of efficient implementation are
addressed in Section~\ref{sec:impl--extens}. Possible extensions of
\textsc{Suffix}-$\sigma$ are sketched in
Section~\ref{sec:extensions}. Our experiments are the subject of
Section~\ref{sec:exper-eval}. In Section~\ref{sec:related-work}, we
put our work into context, before concluding in
Section~\ref{sec:conclusions}.

\section{Preliminaries}
\label{sec:model--background}

We now introduce our model, establish our notation, and provide some
technical background on MapReduce.

\subsection{Data Model}
\label{sec:model}

Our methods operate on sequences of terms (i.e., words or other
textual tokens) drawn from a vocabulary $\mathcal{V}$. We let
$\mathcal{S}$ denote the universe of all sequences over
$\mathcal{V}$. Given a sequence
$\mathbf{s}=\langle\,s_0,\ldots,s_{n-1}\,\rangle$ with
$s_i\in\mathcal{V}$, we refer to its length as $|\mathbf{s}|$, write
$\mathbf{s}[i..j]$ for the subsequence
$\langle\,s_i,\ldots,s_j\,\rangle$, and let $\mathbf{s}[i]$ refer to
the element $s_i$. For two sequences $\mathbf{r}$ and $\mathbf{s}$, we
let $\mathbf{r} \| \mathbf{s}$ denote their concatenation. We say
that
\begin{itemize}
\item $\mathbf{r}$ is a \textit{prefix} of $\mathbf{s}$ ($\mathbf{r} \triangleright \mathbf{s}$)
  iff $$\forall{\,0\le i < |\mathbf{r}|}\::\:\mathbf{r}[i] =
  \mathbf{s}[i]$$
\item $\mathbf{r}$ is a \textit{suffix} of $\mathbf{s}$ ($\mathbf{r} \triangleleft \mathbf{s}$)
  iff $$\forall{\,0\le i <
    |\mathbf{r}|}\::\:\mathbf{r}[i] = \mathbf{s}[|\mathbf{s}| - |\mathbf{r}| + i]$$
\item $\mathbf{r}$ is a \textit{subsequence} of $\mathbf{s}$ ($\mathbf{r} \diamond \mathbf{s}$) iff
  $$\exists{\,0\le j < |\mathbf{s}|} \::\: \forall{\,0\le i <
    |\mathbf{r}|}\::\:\mathbf{r}[i] = \mathbf{s}[i + j]$$
\end{itemize}
and capture how often $\mathbf{r}$ occurs in $\mathbf{s}$ as
$$f(\mathbf{r},\mathbf{s}) = \left|\left\{\, 0 \le j < |\mathbf{s}| \:\:|\:\: \forall{\,0\le i <
      |\mathbf{r}|}\::\:\mathbf{r}[i] = \mathbf{s}[i + j]
    \,\right\}\right|\;.$$

To avoid confusion, we use the following convention: When referring to
sequences of terms having a specific length $k$, we will use the
notion $k$-gram or indicate the considered length by alluding to, for
instance, $5$-grams. The notion $n$-gram, as found in the title, will
be used when referring to variable-length sequences of terms.

As an input, all methods considered in this work receive a document
collection $\mathcal{D}$ consisting of sequences of terms as
documents. Our focus is on determining how often $n$-grams occur in
the document collection. Formally, the \textit{collection frequency}
of an $n$-gram $\mathbf{s}$ is defined as as $cf(\mathbf{s}) =
\sum_{\mathbf{d}\in\mathcal{D}} f(\mathbf{s},\mathbf{d})\;.$
Alternatively, one could consider the document frequency of $n$-grams
as the total number of documents that contain a specific
$n$-gram. While this corresponds to the notion of \textit{support}
typically used in frequent sequence mining, it is less common for
natural language applications. However, all methods presented below
can easily be modified to produce document frequencies instead.

\subsection{MapReduce}
\label{sec:mapreduce}

MapReduce, as described by Dean and Ghemawat~\cite{Dean:2004uq}, is a
programming model and an associated runtime system at Google. While
originally proprietary, the MapReduce programming model has been
widely adopted in practice and several implementations exist. In this
work, we rely on Hadoop~\cite{hadoop} as a popular open-source
MapReduce platform. The objective of MapReduce is to facilitate
distributed data processing on large-scale clusters of commodity
computers. MapReduce enforces a functional style of programming and
lets users express their tasks as two functions
\begin{center}
  \small
  \begin{tabular}{ccc}
    \texttt{map()} & : & \texttt{(k1,v1) -> list<(k2,v2)>}\\
    \texttt{reduce()} & : & \texttt{(k2, list<v2>) -> list<(k3,v3)>}
  \end{tabular}
\end{center}
that consume and emit key-value pairs. Between the \texttt{map}- and
\texttt{reduce}-phase, the system sorts and groups the key-value pairs
emitted by the \texttt{map}-function. The partitioning of key-value
pairs (i.e., how they are assigned to cluster nodes) and their sort
order (i.e., in which order they are seen by the
\texttt{reduce}-function on each cluster node) can be customized, if
needed for the task at hand. For detailed introductions to working
with MapReduce and Hadoop, we refer to Lin and Dyer~\cite{Lin:2010kx}
as well as White~\cite{White:2010ys}.

\section{Methods based on prior ideas}
\label{sec:methods-based-prior}

With our notation established, we next describe three methods based on
prior ideas to compute $n$-gram statistics in MapReduce. Before
delving into their details, let us state the problem that we address
in more formal terms:

\vspace{2mm} 

\textit{Given a document collection $\mathcal{D}$, a minimum
  collection frequency $\tau$, a maximum length $\sigma$, our
  objective is to identify all $n$-grams $\mathbf{s}$ with their
  collection frequency $cf(\mathbf{s})$, for which $cf(\mathbf{s}) \ge
  \tau$ and $|\mathbf{s}| \le \sigma$ hold.}

\vspace{2mm}

We thus assume that $n$-grams are only of interest to the task at
hand, if they occur at least $\tau$ times in the document collection,
coined \textit{frequent} in the following, and consist of at most
$\sigma$ terms. Consider, as an example task, the construction of
$n$-gram language models~\cite{Zhai:2008fk}, for which one would only
look at $n$-grams up to a specific length and/or resort to back-off
models~\cite{Katz:1987fk} to obtain more robust estimates for
$n$-grams that occur less than specific number of times.

The problem statement above can be seen as a special case of frequent
sequence mining that considers only contiguous sequences of
single-element itemsets. We believe this to be an important special
case that warrants individual attention and allows for an efficient
solution in MapReduce, as we show in this work. A more elaborate
comparison to existing research on frequent sequence mining is part of
Section~\ref{sec:related-work}.

To ease our explanations below, we use the following running example,
considering a collection of three documents:
\begin{center}
\begin{tabular}{rcl}
  $\mathbf{d}_1$ & = & \seq{a x b x x}\\
  $\mathbf{d}_2$ & = & \seq{b a x b x}\\
  $\mathbf{d}_3$ & = & \seq{x b a x b}
\end{tabular}
\end{center}
With parameters $\tau=3$ and $\sigma=3$, we expect as output
\begin{center}
\begin{tabular}{rccrccrcc}
  \seq{a} & : & $3$ &  \seq{b} & : & $5$ &  \seq{x} & : & $7$ \\
  \seq{a x} & : & $3$ &  \seq{x b} & : & $4$ & & & \\
  \seq{a x b} & : & $3$ & & & & & & 
\end{tabular}
\end{center}
from any method, when applied to this document collection.

\subsection{Na\"ive Counting}
\label{sec:naive-counting}

One of the example applications of MapReduce, given by Dean and
Ghemawat~\cite{Dean:2004uq} and also used in many tutorials, is word
counting, i.e., determining the collection frequency of every word in
the document collection. It is straightforward to adapt word counting
to consider variable-length $n$-grams instead of only unigrams and
discard those that occur less than $\tau$ times. Pseudo code of this
method, which we coin \textsc{Na\"ive}, is given in
Algorithm~\ref{alg:naive}.

\begin{algorithm}[t]
  \footnotesize
  \SetKwFunction{Emit}{emit} %
  \SetKwFunction{Map}{map} %
  \SetKwFunction{Reduce}{reduce} %

  \texttt{// Mapper}

  \setcounter{AlgoLine}{0}
  \ShowLn\Map{\texttt{long} $did$, \texttt{seq} $\mathbf{d}$} %
  \Begin{ 

    \ShowLn\For{$b=0$ \KwTo $|\mathbf{d}| - 1$}{ %
      \ShowLn\For{$e=b$ \KwTo $min(b+ \sigma - 1, |\mathbf{d}| - 1)$}{ %
        \ShowLn\Emit{\texttt{seq} $\mathbf{d}[b..e]$, \texttt{long} $did$} } %
    } %
  }
  
  \BlankLine

  \texttt{// Reducer}

  \setcounter{AlgoLine}{0}
  \ShowLn\Reduce{\texttt{seq} $\mathbf{s}$, \texttt{list<long>} $l$}
  \Begin{

    \ShowLn\If{$|l| \ge \tau$}{ %

      \ShowLn\Emit{\texttt{seq} $\mathbf{s}$, \texttt{int} $|l|$} %

    } %
  }%

  \caption{\textsc{Na\"ive}}
  \label{alg:naive}
\end{algorithm}

In the \texttt{map}-function, the method emits all $n$-grams of
length up to $\sigma$ for a document together with the document
identifier. If an $n$-gram occurs more than once, it is emitted
multiple times. In the \texttt{reduce}-phase, the collection frequency
of every $n$-gram is determined and, if it exceeds $\tau$, emitted
together with the $n$-gram itself.

Interestingly, apart from minor optimizations, this is the method that
Brants et al.~\cite{Brants:2007ys} used for training large-scale
language models at Google, considering $n$-grams up to length five. In
practice, several tweaks can be applied to improve this simple method
including local pre-aggregation in the \texttt{map}-phase (e.g., using
a combiner in Ha\-doop). Implementation details of this kind are
covered in more detail in Section~\ref{sec:impl--extens}. The
potentially vast number of emitted key-value pairs that needs to be
transferred and sorted, though, remains a shortcoming.

In the worst case, when $\sigma \ge |\mathbf{d}|$, \textsc{Na\"ive}
emits $\mathcal{O}(|\mathbf{d}|^2)$ key-value pairs for a document
$\mathbf{d}$, each consuming $\mathcal{O}(|\mathbf{d}|)$ bytes, so
that the method transfers $\mathcal{O}(|\mathbf{d}|^3)$ bytes between
the \texttt{map}- and \texttt{reduce}-phase. Complementary to that, we
can determine the number of key-value pairs emitted based on the
$n$-gram statistics. \textsc{Na\"ive} emits a total of
$\sum_{\mathbf{s}\in\mathcal{S} : |\mathbf{s}| \le \sigma}
cf(\mathbf{s})$ key-value pairs, each of which consumes
$\mathcal{O}(|\mathbf{s}|)$ bytes.

\subsection{Apriori-Based Methods}
\label{sec:apri-based-meth}

How can one do better than the na\"ive method just outlined? One idea
is to exploit the \textsc{Apriori} principle, as described by Agrawal et
al.~\cite{Agrawal:1994fk} in their seminal paper on identifying
frequent itemsets and follow-up work on frequent pattern
mining~\cite{Agrawal:1995ve,Pei:2004kx,Srikant:1996fk,Zaki:2001ly}. Cast
into our setting, the \textsc{Apriori} principle states that
$$
\mathbf{r} \diamond \mathbf{s} \quad\Rightarrow\quad cf(\mathbf{r}) \ge
cf(\mathbf{s})
$$
holds for any two sequences $\mathbf{r}$ and $\mathbf{s}$, i.e., the
collection frequency of a sequence $\mathbf{r}$ is an upper bound for
the collection frequency of any supersequence $\mathbf{s}$. In what
follows, we describe two methods that make use of the \textsc{Apriori}
principle to compute $n$-gram statistics in MapReduce.

\subsubsection*{\bf\textsc{Apriori-Scan}}

The first \textsc{Apriori}-based method \textsc{Apriori-Scan}, like
the original \textsc{Apriori} algorithm~\cite{Agrawal:1994fk} and
GSP~\cite{Srikant:1996fk}, performs multiple scans over the input
data. During the $k$-th scan the method determines $k$-grams that
occur at least $\tau$ times in the document collection. To this end,
it exploits the output from the previous scan via the \textsc{Apriori}
principle to prune the considered $k$-grams. In the $k$-th scan, only
those $k$-grams are considered whose two constituent $(k-1)$-grams are
known to be frequent. Unlike GSP, that first generates all potentially
frequent sequences as candidates, \textsc{Apriori-Scan} considers only
sequences that actually occur in the document collection. The method
terminates after $\sigma$ scans or when a scan does not produce any
output.

Algorithm~\ref{alg:apriori-scan} shows how the method can be
implemented in MapReduce. The outer \texttt{repeat}-loop controls the
execution of multiple MapReduce jobs, each of which performs one
distributed parallel scan over the input data. In the $k$-th
iteration, and thus the $k$-th scan of the input data, the method
considers all $k$-grams from an input document in the
\texttt{map}-function, but discards those that have a constituent
$(k-1)$-gram that is known to be infrequent. This pruning is done,
leveraging the output from the previous iteration that is kept in a
dictionary. In the \texttt{reduce}-function, analogous to
\textsc{Na\"ive}, collection frequencies of $k$-grams are determined
and output if above the minimum collection frequency $\tau$. After
$\sigma$ iterations or once an iteration does not produce any output,
the method terminates, which is safe since the \textsc{Apriori}
principle guarantees that no longer $n$-gram can occur $\tau$ or more
times in the document collection.

\begin{algorithm}[t]
  \footnotesize

  \SetKwFunction{Emit}{emit} %
  \SetKwFunction{Map}{map} %
  \SetKwFunction{Reduce}{reduce} %
  \SetKwFunction{Get}{get} %
  \SetKwFunction{Add}{add} %
  \SetKwFunction{Keys}{keys} %
  \SetKwFunction{Len}{len} %
  \SetKwFunction{Load}{load} %
  \SetKwFunction{IsEmpty}{isEmpty} %
  \SetKwFunction{Contains}{contains} %
  \SetKwFunction{main}{main} %

  \texttt{int} $k = 1$

  \Repeat{\IsEmpty{output-$(k-1)$} $\vee$ $k = \sigma + 1$}{
    
    \texttt{hashset<int[]>} $dict = $ \Load{output-$(k-1)$}
    
    \BlankLine
    
    \texttt{// Mapper}

    \setcounter{AlgoLine}{0}
    
    \ShowLn\Map{\texttt{long} $did$, \texttt{seq} $\mathbf{d}$} %
    \Begin{     
        
        \ShowLn\For{$b=0$ \KwTo $|\mathbf{d}| - k$}{ %

          \ShowLn\If{$k=1$ $\vee$ \\
            \ShowLn$($\Contains{dict, $\mathbf{d}[b..(b+k-2)]$} $\wedge$\\ \ShowLn~\Contains{dict, $\mathbf{d}[(b+1)..(b+k-1)]$$)$}}{
            \ShowLn\Emit{\texttt{seq} $\mathbf{d}[b..(b+k-1)]$, \texttt{long} $did$}
          }

        } %
              
      } %
    
      \BlankLine
    
      \texttt{// Reducer}

      \setcounter{AlgoLine}{0}
    
      \ShowLn\Reduce{\texttt{seq} $\mathbf{s}$, \texttt{list<long>}
        $l$}
    \Begin{
      
      \ShowLn\If{$|l| \ge \tau$}{ %
        
        \ShowLn\Emit{\texttt{seq} $\mathbf{s}$, \texttt{int} $|l|$} %
        
      } %
    }%
    
    \BlankLine
    
    \setcounter{AlgoLine}{0}
    
    $k$ += $1$

    \BlankLine
   
  } %

  \caption{\textsc{Apriori-Scan}}
  \label{alg:apriori-scan}
\end{algorithm}

When applied to our running example, in its third scan of the input
data, \textsc{Apriori-Scan} emits in the \texttt{map}-phase for every
document $\mathbf{d}_i$ only the key-value pair
$($\seq{a~x~b}$, \mathbf{d}_i)$, but
discards other trigrams (e.g., \seq{b~x~x}) that contain an
infrequent bigram (e.g., \seq{x~x}).

When implemented in MapReduce, every iteration corresponds to a
separate job that needs to be run and comes with its administrative
fix cost (e.g., for launching and finalizing the job). Another
challenge in \textsc{Apriori-Scan} is the implementation of the
dictionary that makes the output from the previous iteration available
and accessible to cluster nodes. This dictionary can either be
implemented locally, so that every cluster node receives a replica of
the previous iteration's output (e.g., implemented using the
distributed cache in Hadoop), or, by loading the output from the
previous iteration into a shared dictionary (e.g., implemented using a
distributed key-value store) that can then be accessed remotely by
cluster nodes. Either way, to make lookups in the dictionary
efficient, significant main memory at cluster nodes is required.

An apparent shortcoming of \textsc{Apriori-Scan} is that it has to
scan the entire input data in every iteration. Thus, although
typically only few frequent $n$-grams are found in later iterations,
the cost of an iteration depends on the size of the input data. The
number of iterations needed, on the other hand, is determined by the
parameter $\sigma$ or the length of the longest frequent $n$-gram.

In the worst case, when $\sigma \ge |\mathbf{d}|$ and $cf(\mathbf{d})
\ge \tau$, \textsc{Apriori-Scan} emits $\mathcal{O}(|\mathbf{d}|^2)$
key-value pairs per document $\mathbf{d}$, each consuming
$\mathcal{O}(|\mathbf{d}|)$ bytes, so that the method transfers
$\mathcal{O}(|\mathbf{d}|^3)$ bytes between the \texttt{map}- and
\texttt{reduce}-phase. Again, we provide a complementary analysis
based on the actual $n$-gram statistics. To this end, let
$$
\mathcal{S}_{NP} = \left\{ \mathbf{s} \in \mathcal{S} \:|\:
  \forall{\,\mathbf{r} \in \mathcal{S}} \::\: (\mathbf{r} \neq
  \mathbf{s} \:\wedge\: \mathbf{r} \diamond \mathbf{s}) \Rightarrow
  cf(\mathbf{r}) \ge \tau \right\}
$$
denote the set of sequences that cannot be pruned based on the
\textsc{Apriori} principle, i.e., whose true subsequences all occur at
least $\tau$ times in the document collection. \textsc{Apriori-Scan}
emits a total of $\sum_{\mathbf{s} \in \mathcal{S}_{NP} : |\mathbf{s}|
  \le \sigma} cf(\mathbf{s})$ key-value pairs, each of which amounts
to $\mathcal{O}(|\mathbf{s}|)$ bytes. Obviously, $\mathcal{S}_{NP}
\subseteq \mathcal{S}$ holds, so that \textsc{Apriori-Scan} emits at
most as many key-value pairs as \textsc{Na\"ive}. Its concrete gains,
though, depend on the value of $\tau$ and characteristics of the
document collection.

\subsubsection*{\bf\textsc{Apriori-Index}}

The second \textsc{Apriori}-based method \textsc{Apriori-Index} does not
repeatedly scan the input data but incrementally builds an inverted
index of frequent $n$-grams from the input data as a more compact
representation. Operating on an index structure as opposed to the
original data and considering $n$-grams of increasing length, it
resembles SPADE~\cite{Zaki:2001ly} when breadth-first traversing the
sequence lattice.

\begin{algorithm}[t]
  \footnotesize

  \SetKwFunction{Emit}{emit} %
  \SetKwFunction{Map}{map} %
  \SetKwFunction{Reduce}{reduce} %
  \SetKwFunction{Get}{get} %
  \SetKwFunction{Add}{add} %
  \SetKwFunction{Keys}{keys} %
  \SetKwFunction{Len}{len} %
  \SetKwFunction{Join}{join} %
  \SetKwFunction{IsEmpty}{isEmpty} %
  \SetKwFunction{CF}{cf} %
  \SetKwFunction{main}{main} %
  
  \texttt{int} $k$ $=$ $1$
  
  \Repeat{\IsEmpty{output-$(k-1)$} $\vee$ $k =$  $min(\sigma, \mathtt{K})$}{

    \If{$k \le \mathtt{K}$}{

      \BlankLine
      
      \texttt{// Mapper \#1}
      
      \setcounter{AlgoLine}{0}
      
      \ShowLn\Map{\texttt{long} $did$, \texttt{seq} $\mathbf{d}$}
      \Begin{     
        
        \ShowLn\texttt{hashmap<seq, int[]>} pos $= \emptyset$
        
        \ShowLn\For{$b=0$ \KwTo $|\mathbf{d}|-1$}{
          
          \ShowLn\Add{\Get{pos, $\mathbf{d}[b..(b+k-1)]$}, $b$}
          
        } %
        
        \ShowLn\For{\texttt{seq} $\mathbf{s}$ : \Keys{pos}}{ %
          \ShowLn\Emit{\hbox{\texttt{seq} $\mathbf{s}$, \texttt{posting} $($$did$, \Get{pos,$\mathbf{s}$}$)$}}
        } %
         
      } %
       
      \BlankLine
       
      \texttt{// Reducer \#1}
      
      \setcounter{AlgoLine}{0}
      
      \ShowLn\Reduce{\texttt{seq}~$\mathbf{s}$,~\texttt{list<posting>}~$l$}
      \Begin{
        
        \ShowLn\If{\CF{$l$} $\ge \tau$}{ %
           
          \ShowLn\Emit{\texttt{seq} $\mathbf{s}$, \texttt{list<posting>} $l$} %
           
        } 
         
      }
      }
      \Else{               
      
      \setcounter{AlgoLine}{0}
      
      \BlankLine
      
      \texttt{// Mapper \#2}
      
      \setcounter{AlgoLine}{0}
      
      \ShowLn\Map{\texttt{seq} $\mathbf{s}$, \texttt{list<posting>} $l$}
      \Begin{
        
        \ShowLn\Emit{\texttt{seq} $\mathbf{s}[0..|\mathbf{s}|-2]$,\\ \ShowLn\quad\quad\quad\texttt{$($r-seq, list<posting>$)$} $(\mathbf{s},l)$}%
        
        \ShowLn\Emit{\texttt{seq} $\mathbf{s}[1..|\mathbf{s}|-1]$,\\
          \ShowLn\quad\quad\quad\texttt{$($l-seq, list<posting>$)$} $(\mathbf{s},l)$}%
        
      }%
      
      \BlankLine
      
      \texttt{// Reducer \#2}
      
      \setcounter{AlgoLine}{0}
      
      \ShowLn\Reduce{\hbox{\texttt{seq} $\mathbf{s}$, \texttt{list<$($seq, list<posting>$)$>} $l$}}
      \Begin{   
        
        \ShowLn\For{$($\texttt{l-seq}, \texttt{list<posting>$)$} $(\mathbf{m},l_m)$ : $l$}{ %
          
          \ShowLn\For{$($\texttt{r-seq}, \texttt{list<posting>$)$} $(\mathbf{n},l_n)$ : $l$}{ %
            
            \ShowLn\texttt{list<posting>} $l_{j}$ = \Join{$l_m$, $l_n$}
           
            \ShowLn\If{\CF{$l_{j}$} $\ge \tau$}{ %
              
              \ShowLn\texttt{seq} $\mathbf{j}$ $= \mathbf{m} \,\|\, \langle\,\mathbf{n}[|\mathbf{n}|-1]\,\rangle$
             
              \ShowLn\Emit{\texttt{seq} $\mathbf{j}$, \texttt{list<posting>} $l_j$}
              
            } %
            
          } %
          
        } %
        
      }%
    }
    
    \BlankLine
    
    $k$ += $1$
    
    \BlankLine
    
  }%
  
  \caption{\textsc{Apriori}-Index}
  \label{alg:apriori-index}
\end{algorithm}

Pseudo code of \textsc{Apriori-Index} is given in
Algorithm~\ref{alg:apriori-index}. In its first phase, the method
constructs an inverted index with positional information for all
frequent $n$-grams up to length $\mathtt{K}$ (cf. \texttt{Mapper~\#1}
and \texttt{Reducer~\#1} in the pseudo code). In its second phase, to
identify frequent $n$-grams beyond that length, \textsc{Apriori-Index}
harnesses the output from the previous iteration. Thus, to determine a
frequent $k$-gram (e.g., \seq{b~a~x}), the method joins the posting
lists of its constituent $(k-1)$-grams (i.e., \seq{b~a} and
\seq{a~x}). In MapReduce, this can be accomplished as follows
(cf. \texttt{Mapper~\#2} and \texttt{Reducer~\#2} in the pseudo code):
The \texttt{map}-function emits for every frequent $(k-1)$-gram two
key-value pairs. The frequent $(k-1)$-gram itself along with its
posting list serves in both as a value. As keys the prefix and suffix
of length $(k-2)$ are used. In the pseudo code, the method keeps track
of whether the key is a prefix or suffix of the sequence in the value
by using the \texttt{r-seq} and \texttt{l-seq} subtypes. The
\texttt{reduce}-function identifies for a specific key all
compatible sequences from the values, joins their posting lists, and
emits the resulting $k$-gram along with its posting list if its
collection frequency is at least $\tau$. Two sequences are compatible
and must be joined, if one has the current key as a prefix, and the
other has it as a suffix. In its nested \texttt{for}-loops, the method
considers all compatible combinations of sequences. This second phase
of \textsc{Apriori-Index} can be seen as a distributed candidate
generation and pruning step.

Applied to our running example and assuming $K=2$, the method only
sees one pair of compatible sequences with their posting lists for the
key \seq{x} in its third iteration,
namely:
\begin{center}
\begin{tabular}{rcl}
  \seq{a x} & : & $\langle\,\mathbf{d}_1:[0],\:\:\mathbf{d}_2:[1],\:\:\mathbf{d}_3:[2]\,\rangle$ \\
  \seq{x b} & : & $\langle\,\mathbf{d}_1:[1],\:\:\mathbf{d}_2:[2],\:\:\mathbf{d}_3:[0,3]\,\rangle\;.$
\end{tabular}
\end{center}
By joining those, \textsc{Apriori-Index} obtains the only frequent
$3$-gram with its posting list
\begin{center}
\begin{tabular}{rcl}
  \seq{a x b} & : & $\langle\,\mathbf{d}_1:[0],\:\:\mathbf{d}_2:[1],\:\:\mathbf{d}_3:[2]\,\rangle\;.$
\end{tabular}
\end{center}

For all $k < \mathtt{K}$, it would be enough to determine only
collection frequencies, as opposed to, positional information of
$n$-grams. While a straightforward optimization in practice, we opted
for simpler pseudo code. When implemented as described in
Algorithm~\ref{alg:apriori-index}, the method produces an inverted
index with positional information that can be used to quickly
determine the locations of a specific frequent $n$-gram.

One challenge when implementing \textsc{Apriori-Index} is that the
number and size of posting-list values seen for a specific key can
become large in practice. Moreover, to join compatible sequences,
these posting lists have to be buffered, and a scalable implementation
must deal with the case when this is not possible in the available
main memory. This can, for instance, be accomplished by storing
posting lists temporarily in a disk-resident key-value store.

The number of iterations needed by \textsc{Apriori-Index} is
determined by the parameter $\sigma$ or the length of the longest
frequent $n$-gram. Since every iteration, as for
\textsc{Apriori-Scan}, corresponds to a separate MapReduce job, a
non-negligible administrative fix cost is incurred.

In the worst case, when $\sigma \ge |\mathbf{d}|$ and $cf(\mathbf{d})
\ge \tau$, \textsc{Apriori-Index} emits $\mathcal{O}(|\mathbf{d}|^2)$
key-value pairs per document $\mathbf{d}$, each consuming
$\mathcal{O}(|\mathbf{d}|)$ bytes, so that
$\mathcal{O}(|\mathbf{d}|^3)$ bytes are transferred the \texttt{map}-
and \texttt{reduce}-phase. We assume $K < \sigma$ for the
complementary analysis. In its first $K$ iterations,
\textsc{Apriori-Index} emits $\sum_{\mathbf{s} \in \mathcal{S} :
  |\mathbf{s}| \le K} df(\mathbf{s})$ key-value pairs, where
$df(\mathbf{s}) \le cf(\mathbf{s})$ refers to the document frequency
of the $n$-gram $\mathbf{s}$, as mentioned in
Section~\ref{sec:model--background}. Each key-value pair consumes
$\mathcal{O}(cf(\mathbf{s}))$ bytes. To analyze the following
iterations, let
$$
\mathcal{S}_{F} = \left\{ \mathbf{s} \in \mathcal{S} \:|\:
  cf(\mathbf{s}) \ge \tau \right\}
$$
denote the set of frequent $n$-grams that occur at least $\tau$
times. \textsc{Apriori-Index} emits a total of
$$
2\cdot|\{\mathbf{s} \in \mathcal{S}_{F} \:|\: K \le |\mathbf{s}| < \sigma \}|
$$
key-value pairs, each of which consumes $\mathcal{O}(cf(\mathbf{s}))$
bytes. Like for \textsc{Apriori-Scan}, the concrete gains depend on
the value of $\tau$ and characteristics of the document collection.\\

\vspace{-4mm}
\section{Suffix sorting \& aggregation}
\label{sec:suffix-sorting}

As already argued, the methods presented so far suffer from either
excessive amounts of data that need to be transferred and sorted,
requiring possibly many MapReduce jobs, or a high demand for main
memory at cluster nodes. Our novel method \textsc{Suffix}-$\sigma$
avoids these deficiencies: It requires a single MapReduce job,
transfers only a modest amount of data, and requires little main
memory at cluster nodes.

Consider again what the \texttt{map}-function in the
\textsc{Na\"ive} approach emits for document $\mathbf{d}_3$ from our
running example. Emitting key-value pairs for all of the $n$-grams
\seq{b~a~x}, \seq{b~a}, and \seq{b} is clearly wasteful. The key
observation here is that the latter two are subsumed by the first one
and can be obtained as its prefixes. Suffix
arrays~\cite{Manber:1993vn} and other string processing techniques
exploit this very idea.

Based on this observation, it is safe to emit key-value pairs only for
a subset of the $n$-grams contained in a document. More precisely, it
is enough to emit at every position in the document a single key-value
pair with the suffix starting at that position as a key. These
suffixes can further be truncated to length $\sigma$ -- hence the name
of our method.

\begin{algorithm}[t]
  \footnotesize

  \SetKwFunction{Emit}{emit} %
  \SetKwFunction{Map}{map} %
  \SetKwFunction{Reduce}{reduce} %
  \SetKwFunction{Cleanup}{cleanup} %
  \SetKwFunction{Partition}{partition} %
  \SetKwFunction{Compare}{compare} %
  \SetKwFunction{Return}{return} %
  \SetKwFunction{Hashcode}{hashcode} %
  \SetKwFunction{LCP}{lcp} %
  \SetKwFunction{Peek}{peek} %
  \SetKwFunction{Push}{push} %
  \SetKwFunction{Pop}{pop} %
  \SetKwFunction{Len}{len} %
  \SetKwFunction{Seq}{seq} %

  \texttt{// Mapper}

  \setcounter{AlgoLine}{0}

  \ShowLn\Map{\texttt{long} $did$, \texttt{seq} $\mathbf{d}$} %
  \Begin{     
    \ShowLn\For{$b=0$ \KwTo $|\mathbf{d}|-1$}{ %
      \ShowLn\Emit{\texttt{seq} $\mathbf{d}[b..min(b + \sigma - 1, |\mathbf{d}|-1)]$, \texttt{long} $did$} %
    } %
  } %
  
  \BlankLine

  \texttt{// Reducer}

  \texttt{stack<int>} $terms = \emptyset$ %

  \texttt{stack<int>} $counts = \emptyset$ %

  \setcounter{AlgoLine}{0}

  \ShowLn\Reduce{\texttt{seq} $\mathbf{s}$, \texttt{list<long>} $l$}
  \Begin{

    \ShowLn\While{\LCP{$\textbf{s}$,\Seq{$terms$}} $<$ \Len{$terms$}}{ %

      \ShowLn\If{\Peek{$counts$} $\ge \tau$ }{%

        \ShowLn\Emit{\texttt{seq} \Seq{$terms$}, \texttt{int} \Peek{$counts$}}

        } %

     \ShowLn\Pop{$terms$} %

     \ShowLn\Push{$counts$, \Pop{$counts$} + \Pop{$counts$}} %

      } %

      \BlankLine

      \ShowLn\eIf{\Len{$terms$} $= |\mathbf{s}|$}{\ShowLn %
        \ShowLn\Push{$counts$, \Pop{$counts$} + $|l|$}
      }{ %
        \ShowLn\For{$i=$ \LCP{$\textbf{s}$, \Seq{$terms$}} \KwTo $|\mathbf{s}| - 1$}{ %
          \ShowLn\Push{$terms$, $\mathbf{s}[i]$}

          \ShowLn\Push{$counts$, $(i == |\mathbf{s}| - 1$ ? $|l|$ : $0)$}
        }
      } %

  }%

  \BlankLine

  \setcounter{AlgoLine}{0}

  \ShowLn\Cleanup{}
  \Begin{
    \ShowLn\Reduce{\texttt{seq} $\emptyset$, \texttt{list<long>} $\emptyset$}
  }%

  \BlankLine

  \texttt{// Partitioner}

  \setcounter{AlgoLine}{0}

  \ShowLn\Partition{\texttt{seq} $\mathbf{s}$}
  \Begin{
    
    \ShowLn\Return \Hashcode{$\mathbf{s}[0]$} \texttt{mod} $R$

  }%

  \BlankLine

  \texttt{// Comparator}

  \setcounter{AlgoLine}{0}

  \ShowLn\Compare{\texttt{seq} $r$, \texttt{seq} $s$}
  \Begin{
    \ShowLn\For{$b=0$ \KwTo $min(|\mathbf{r}|, |\mathbf{s}|) - 1$}{ %
      \ShowLn\uIf{$\mathbf{r}[b] < \mathbf{s}[b]$}{ %
         \ShowLn\Return $+1$
      } %
      \ShowLn\ElseIf{$\mathbf{r}[b] > \mathbf{s}[b]$}{ %
        \ShowLn\Return $-1$
      } %
    } %
    \ShowLn\Return $|\mathbf{s}| - |\mathbf{r}|$ %
  }%

  \caption{\textsc{Suffix}-$\sigma$}
  \label{alg:suffix}
\end{algorithm}

To determine the collection frequency of a specific $n$-gram
$\mathbf{r}$, we have to determine how many of the suffixes emitted in
the \texttt{map}-phase are prefixed by $\mathbf{r}$. To do so
correctly using only a single MapReduce job, we must ensure that all
relevant suffixes are seen by the same reducer. This can be
accomplished by partitioning suffixes based on their first term only,
as opposed to, all terms therein. It is thus guaranteed that a single
reducer receives all suffixes that begin with the same term. This
reducer is then responsible for determining the collection frequencies
of all $n$-grams starting with that term. One way to accomplish this
would be to enumerate all prefixes of a received suffix and aggregate
their collection frequencies in main memory (e.g., using a hashmap or
a prefix tree). Since it is unknown whether an $n$-gram is represented
by other yet unseen suffixes from the input, it cannot be emitted
early along with its collection frequency. Bookkeeping is thus needed
for many $n$-grams and requires significant main memory.

How can we reduce the main-memory footprint and emit $n$-grams with
their collection frequency early on? The key idea is to exploit that
the order in which key-value pairs are sorted and received by reducers
can be influenced. \textsc{Suffix}-$\sigma$ sorts key-value pairs in
reverse lexicographic order of their suffix key, formally defined as
follows for sequences $\mathbf{r}$ and $\mathbf{s}$:
\begin{eqnarray}
\mathbf{r} < \mathbf{s} \Leftrightarrow (|\mathbf{r}| > |\mathbf{s}| \:\wedge\: \mathbf{s} \triangleright \mathbf{r}) \:\vee\: \hspace{4.4cm}\nonumber\\
\exists{\,0 \!\le\! i < min(|\mathbf{r}|, |\mathbf{s}|)} \!:\! \mathbf{r}[i] > \mathbf{s}[i] \wedge \forall{\,0 \!\le\! j < i} : \mathbf{r}[j] = \mathbf{s}[j]\;. \nonumber 
\end{eqnarray}

To see why this is useful, recall that each suffix from the input
represents all $n$-grams that can be obtained as its prefixes. Let
$\mathbf{s}$ denote the current suffix from the input. The reverse
lexicographic order guarantees that we can safely emit any $n$-gram
$\mathbf{r}$ such that $\mathbf{r} < \mathbf{s}$, since no yet unseen
suffix from the input can represent $\mathbf{r}$. Conversely, at this
point, the only $n$-grams for which we have to do bookkeeping, since
they are represented both by the current suffix $\mathbf{s}$ and
potentially by yet unseen suffixes, are the prefixes of
$\mathbf{s}$. We illustrate this observation with our running
example. The reducer responsible for suffixes starting with \term{b}
receives:
\begin{center}
\begin{tabular}{lcl}
  \seq{b x x} 	& : & $\langle\,\mathbf{d}_1\,\rangle$\\
  \seq{b x}    	 & : &  $\langle\,\mathbf{d}_2\,\rangle$\\
  \seq{b a x}  	 & : &  $\langle\,\mathbf{d}_2,\:\:\mathbf{d}_3\,\rangle$\\
  \seq{b}	 & : & $\langle\,\mathbf{d}_3\,\rangle\;.$
\end{tabular}
\end{center}
When seeing the third suffix \seq{b~a~x}, we can immediately finalize
the collection frequency of the $n$-gram \seq{b~x} and emit it, since
no yet unseen suffix can have it as a prefix. On the contrary, the
$n$-grams \seq{b} and \seq{b~a} cannot be emitted, since yet unseen
suffixes from the input may have them as a prefix.

\begin{figure}[h]
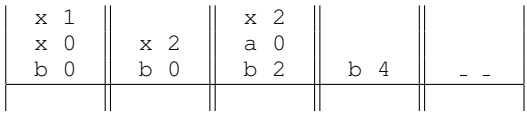

  \centering  
  {\small\tt
  \begin{tabular}{|c||c||c||c||c|}
   x 1 &     & x 2 &     &      \\
   x 0 & x 2 & a 0 &     &      \\
   b 0 & b 0 & b 2 & b 4 & \_ \_ \\\hline
   \hspace{9mm} & \hspace{9mm} & \hspace{9mm} & \hspace{9mm} & \hspace{9mm} \\   
  \end{tabular}
  }
  \caption{\textsc{Suffix}-$\sigma$'s bookkeeping illustrated}
  \label{fig:SuffixBookkeeping}
\end{figure}

Building on this observation, we can do efficient bookkeeping for
prefixes of the current suffix $\mathbf{s}$ only and lazily aggregate
their collection frequencies using two stacks. On the first stack
$terms$, we keep the terms constituting $\mathbf{s}$. The~second stack
$counts$ keeps one counter per prefix of $\mathbf{s}$. Between
invocations of the \texttt{reduce}-function, we maintain two
invariants. First, the two stacks have the same size $m$. Second,
$\sum_{j=i}^{m-1} counts[j]$ reflects how often the $n$-gram
$\langle\,terms[0],\,\ldots,\,terms[i]\,\rangle$ has been seen so far
in the input. To maintain these invariants, when processing a suffix
$\mathbf{s}$ from the input, we first synchronously pop elements from
both stacks until the contents of $terms$ form a prefix of
$\mathbf{s}$. Before each pop operation, we emit the contents of
$terms$ and the top element of $counts$, if the latter is above our
minimum collection frequency $\tau$. When popping an element from
$counts$, its value is added to the new top element. Following that,
we update $terms$, so that its contents equal the suffix
$\mathbf{s}$. For all but the last term added, a zero is put on
$counts$. For the last term, we put the frequency of $\mathbf{s}$,
reflected by the length of its associated document-identifier list
value, on $counts$. Figure~\ref{fig:SuffixBookkeeping} illustrates how
the states of the two stacks evolve, as the above example input is
processed.

Pseudo code of \textsc{Suffix}-$\sigma$ is given in
Algorithm~\ref{alg:suffix}. The \texttt{map}-function emits for
every document all its suffixes truncated to length $\sigma$ if
possible. The \texttt{reduce}-function reads suffixes in reverse
lexicographic order and performs the bookkeeping using two separate
stacks for $n$-grams (\textit{terms}) and their collection frequencies
(\textit{counts}), as described above. The function \texttt{seq()}
returns the $n$-gram corresponding to the entire \textit{terms}
stack. The function \texttt{lcp()} returns the length of the longest
common prefix that two $n$-grams share. In addition,
Algorithm~\ref{alg:suffix} contains a \texttt{partition}-function
ensuring that suffixes are assigned to one of $R$ reducers solely
based on their first term, as well as, a \texttt{compare}-function
that ensures the reverse lexicographic order of input suffixes in the
\texttt{map}-phase. When implemented in Hadoop, these two functions
would materialize as a custom partitioner class and a custom
comparator class. Finally, \texttt{cleanup()} is a method invoked
once, when all input has been seen.

\textsc{Suffix}-$\sigma$ emits $\mathcal{O}(|\mathbf{d}|)$ key-value
pairs per document $\mathbf{d}$. Each of these key-value pairs
consumes $\mathcal{O}(|\mathbf{d}|)$ bytes in the worst case when
$\sigma \ge |\mathbf{d}|$. The method thus transfer
$\mathcal{O}(|\mathbf{d}|^2)$ bytes between the \texttt{map}- and
\texttt{reduce}-phase. For every term occurrence in the document
collection, \textsc{Suffix}-$\sigma$ emits exactly one key-value pair,
so that in total $\sum_{\mathbf{s} \in \mathcal{S} : |\mathbf{s}| = 1}
cf(\mathbf{s})$ key-value pairs are emitted, each consuming
$\mathcal{O}(\sigma)$ bytes. 

\vspace{-2mm}
\section{Efficient implementation}
\label{sec:impl--extens}
Having described the different methods at a conceptual level, we now
provide details on aspects of their implementation, which we found to
have a significant impact on performance in practice:

\textbf{Document Splits.} Collection frequencies of individual terms
(i.e., unigrams) can be exploited to drastically reduce required work
by splitting up every document at infrequent terms that it
contains. Thus, assuming that \term{z} is an infrequent term given
the current value of $\tau$, we can split up a document like \seq{c b
  a z b a c} into the two shorter sequences \seq{c b a} and \seq{b a
  c}. Again, this is safe due to the \textsc{Apriori} principle, since
no frequent $n$-gram can contain~\term{z}. All methods profit from
this -- for large values of $\sigma$ in particular.

\textbf{Sequence Encoding.} It is inefficient to operate on documents
in a textual representation. As a one-time preprocessing, we therefore
convert our document collections, so that they are represented as a
dictionary, mapping terms to term identifiers, and one integer
term-identifier sequence for every document. We assign identifiers to
terms in descending order of their collection frequency to optimize
compression. From there on, our implementation internally only deals
with arrays of integers. Whenever serialized for transmission or
storage, these are compactly represented using variable-byte
encoding~\cite{Witten:1999fk}. This also speeds up sorting, since
$n$-grams can now be compared using integer operations as opposed to
operations on strings, thus requiring generally fewer machine
instructions. Compact sequence encoding benefits all methods -- in
particular \textsc{Apriori-Scan} with its repeated scans
of the document collection.

\textbf{Key-Value Store.} For \textsc{Apriori-Scan} and
\textsc{Apriori-Index}, reducers potentially buffer a lot of data,
namely, the dictionary of frequent $(k-1)$-grams or the set of posting
lists to be joined. Our implementation keeps this data in main memory
as long as possible. Otherwise, it migrates the data into a
disk-resident key-value store (Berkeley DB Java
Edition~\cite{berkeleydb}). Most main memory is then used for caching,
which helps \textsc{Apriori-Scan} in particular, since lookups of
frequent $(k-1)$-grams typically hit the cache.

\textbf{Hadoop-Specific Optimizations} that we use in our
implementation include local aggregation (cf. \texttt{Mapper \#1} in
Algorithm~\ref{alg:apriori-index}), Hadoop's distributed cache
facility, raw comparators to avoid deserialization and object
instantiation, as well as other best practices (e.g., described
in~\cite{White:2010ys}).

How easy to implement are the methods presented in previous sections?
While hard to evaluate systematically, we still want to address this
question based on our own experience. \textsc{Na\"ive} is the clear
winner here. Implementations of the \textsc{Apriori}-based methods, as
explained in Section~\ref{sec:methods-based-prior}, require various
tweaks (e.g., the use of a key-value store) to make them
work. \textsc{Suffix}-$\sigma$ does not require any of those and, when
Hadoop is used as a MapReduce implementation, can be implemented using
only on-board functionality.

\vspace{-1mm}
\section{Extensions}
\label{sec:extensions}

In this section, we describe how \textsc{Suffix}-$\sigma$ can be
extended to consider only maximal/closed $n$-grams and thus produce a
more compact result. Moreover, we explain how it can support
aggregations beyond occurrence counting, using $n$-gram time series,
recently made popular by~\cite{Michel:ys}, as an example.

\vspace{-1mm}
\subsection{Maximality \& Closedness}
\label{sec:clos--maxim}
\vspace{-1mm}

The number of $n$-grams that occur at least $\tau$ times in the
document collection can be huge in practice. To reduce it, we can
adopt the notions of maximality and closedness common in frequent
pattern mining. Formally, an $n$-gram $\mathbf{r}$ is
\textit{maximal}, if there is no $n$-gram $\mathbf{s}$ such that
$\mathbf{r} \diamond \mathbf{s}$ and $cf(\mathbf{s}) \ge
\tau$. Similarly, an $n$-gram $\mathbf{r}$ is \textit{closed}, if no
$n$-gram $\mathbf{s}$ exists such that $\mathbf{r} \diamond
\mathbf{s}$ and $cf(\mathbf{r}) = cf(\mathbf{s}) \ge \tau$. The sets
of maximal or closed $n$-grams are subsets of all $n$-grams that occur
at least $\tau$ times. Omitted $n$-grams can be reconstructed -- for
closedness even with their accurate collection frequency.

\textsc{Suffix}-$\sigma$ can be extended to produce maximal or closed
$n$-grams. Recall that, in its \texttt{reduce}-function, our method
processes suffixes in reverse lexicographic order. Let $\mathbf{r}$
denote the last $n$-gram emitted. For maximality, we only emit the
next $n$-gram $\mathbf{s}$, if it is no prefix of $\mathbf{r}$ (i.e.,
$\neg(\mathbf{s} \triangleright \mathbf{r})$). For closedness, we only
emit $\mathbf{s}$, if it is no prefix of $\mathbf{r}$ or if it has a
different collection frequency (i.e., $\neg(\mathbf{s} \triangleright
\mathbf{r} \:\wedge\: cf(\mathbf{s}) = cf(\mathbf{r}))$). In our
example, the reducer responsible for term \term{a} receives
\begin{center}
\begin{tabular}{lcl}
  \seq{a x b} & : & $\langle\,\mathbf{d}_1,\:\:\mathbf{d}_2,\:\:\mathbf{d}_3\,\rangle$
\end{tabular}
\end{center}
and, both for maximality and closedness, emits only the $n$-gram
\seq{a x b} but none of its prefixes. With this extension, we thus
emit only prefix-maximal or prefix-closed $n$-grams, whose formal
definitions are analogous to those of maximality and closedness above,
but replace $\diamond$ by $\triangleright$. In our example, we still
emit \seq{x b} and \seq{b} on the reducers responsible for terms
\term{x} and \term{b}, respectively. For maximality, as subsequences
of \seq{a x b}, these $n$-grams must be omitted. We achieve this by
means of an additional post-filtering MapReduce job. As input, the job
consumes the output produced by \textsc{Suffix}-$\sigma$ with the
above extensions. In its \texttt{map}-function, $n$-grams are
reversed (e.g., \seq{a x b} becomes \seq{b x a}). These reversed
$n$-grams are partitioned based on their first term and sorted in
reverse lexicographic order, reusing ideas from
\textsc{Suffix}-$\sigma$. In the \texttt{reduce}-function, we apply
the same filtering as described above to keep only prefix-maximal or
prefix-closed reversed $n$-grams. Before emitting a reversed $n$-gram,
we restore its original order by reversing it. In our example, the
reducer responsible for \term{b} receives
\begin{center}
\begin{tabular}{lcl}
  \seq{b x a} 	& : & $3$\\
  \seq{b x}   	& : & $4$\\
  \seq{b}	& : & $5$\\
\end{tabular}
\end{center}
and, for maximality, only emits \seq{a x b}. In summary, we obtain
maximal or closed $n$-grams by first determining prefix-maximal or
prefix-closed $n$-grams and, after that, identifying the
suffix-maximal or suffix-closed among them.

\vspace{-1mm}
\subsection{Beyond Occurrence Counting}
\label{sec:aggr-beyond-count}
\vspace{-1mm}

Our focus so far has been on determining collection frequencies of
$n$-grams, i.e., counting their occurrences in the document
collection. One can move beyond occurrence counting and aggregate
other information about $n$-grams, e.g.:
\begin{itemize}

\item build an \textit{inverted index} that records for every $n$-gram
  how often or where it occurs in individual documents;

\item compute \textit{statistics based on meta-data of documents}
  (e.g., timestamp or location) that contain a $n$-gram.

\end{itemize}
In the following, we concentrate on the second type of aggregation
and, as a concrete instance, consider the computation of $n$-gram time
series. Here, the objective is to determine for every $n$-gram a time
series whose observations reveal how often the $n$-gram occurs in
documents published, e.g., in a specific
year. \textsc{Suffix}-$\sigma$ can be extended to produce such
$n$-gram time series as follows: In the \texttt{map}-function we
emit every suffix along with the document identifier and its
associated timestamp. In the \texttt{reduce}-function, the
\textit{counts} stack is replaced by a stack of time series, which we
aggregate lazily. When popping an element from the stack, instead of
adding counts, we add time series observations. In the same manner, we
can compute other statistics based on the occurrences of an $n$-gram
in documents and their associated meta-data. While these could also be
computed by an extension of \textsc{Na\"ive}, the benefit of using
\textsc{Suffix}-$\sigma$ is that the required document meta-data is
transferred only per suffix of a document, as opposed to, per
contained $n$-gram.

\vspace{-1mm}
\section{Experimental evaluation}
\label{sec:exper-eval}

We conducted comprehensive experiments to compare the different
methods and understand their relative benefits and trade-offs. Our
findings from these experiments are the subject of this section.

\vspace{-1mm}
\subsection{Setup \& Implementation}
\label{sec:setup--datasets}
\vspace{-1mm}

\textbf{Cluster Setup.} All experiments were run on a local cluster
consisting of ten Dell R410 server-class computers, each equipped with
64~GB of main memory, two Intel Xeon X5650 6-core CPUs, and four
internal 2~TB SAS 7,200 rpm hard disks configured as a
bunch-of-disks. Debian GNU/Linux 5.0.9 (Lenny) was used as an
operating system. Machines in the cluster are connected via 1~GBit
Ethernet. We use Cloudera CDH3u0 as a distribution of Hadoop~0.20.2
running on Oracle Java 1.6.0\_26. One of the machines acts a master
and runs Hadoop's namenode and jobtracker; the other nine machines are
configured to run up to ten map tasks and ten reduce tasks in
parallel. To restrict the number of map/reduce slots, we employ a
capacity-constrained scheduler pool in Hadoop. When we state that $n$
map/reduce slots are used, our cluster executes up to $n$~map tasks
and $n$~reduce tasks in parallel. Java virtual machines to process
tasks are always launched with 4~GB heap space.

\textbf{Implementation.} All methods are implemented in Java (JDK 1.6)
applying the optimizations described in Section~\ref{sec:impl--extens}
to the extent possible and sensible for each of them.

\textbf{Methods.} We compare the methods \textsc{Na\"ive},
\textsc{Apriori-Scan}, \textsc{Apriori-Index}, and
\textsc{Suffix}-$\sigma$ in our experiments. For
\textsc{Apriori-Index}, we set $K=4$, so that the method directly
computes collection frequencies of $n$-grams having length four or
less. We found this to be the best-performing parameter setting in a
series of calibration experiments.

\textbf{Measures.} For our experiments in the following, we report as
performance measures:
\begin{itemize}
\item[(a)] \textit{wallclock time} as the total time elapsed between
  launching a method and receiving the final result (possibly
  involving multiple Hadoop jobs),

\item[(b)] \textit{bytes transferred} as the total amount of data
  transferred between \texttt{map}- and \texttt{reduce}-phase(s)
  (obtained from Hadoop's \smalltt{MAP\_OUTPUT\_BYTES} counter),

\item[(c)] \textit{\# records} as the total number of key-value pairs
  transferred and sorted between \texttt{map}- and
  \texttt{reduce}-phase(s) (obtained from Hadoop's
  \smalltt{MAP\_OUTPUT\_RECORDS} counter).
\end{itemize}
For \textsc{Apriori-Scan} and \textsc{Apriori-Index}, measures (b) and
(c) are aggregates over all Hadoop jobs launched. All measurements
reported are based on single runs and were performed with exclusive
access to the Hadoop cluster, i.e., without concurrent activity by
other jobs, services, or users.

\begin{table}[t]
  \centering
  \caption{Dataset characteristics}
  \label{tab:Datasets}
  \begin{tabular}[h]{lrr}
  \toprule
    				& \multicolumn{1}{l}{\textbf{NYT}} 	& \multicolumn{1}{l}{\textbf{C09}} \\
  \midrule                                
  \# documents  		& $1,830,592$  		& $50,221,915$	\\
  \# term occurrence		& $1,049,440,645$	& $21,404,321,682$	\\
  \# distinct terms		& $345,827$		& $979,935$ \\
  \# sentences			& $55,362,552$		& $1,257,357,167$	\\
  sentence length (mean)\!\!\!\!\!\!	& $18.96$		& $17.02$	\\
  sentence length (stddev)\!\!\!\!\!\!	& $14.05$		& $17.56$	\\
  \bottomrule
  \end{tabular}
\end{table}

\vspace{-1mm}
\subsection{Datasets}
\label{sec:datasets}
\vspace{-1mm}

We use two publicly-available real-world datasets for our experiments,
namely:

\begin{itemize}

\item \textbf{The New York Times Annotated Corpus}~\cite{nyt}
  consisting of more than 1.8 million newspaper articles from the
  period 1987--2007 (NYT);

\item \textbf{ClueWeb09-B}~\cite{cw}, as a well-defined subset of the
  ClueWeb09 corpus of web documents, consisting of more than 50
  million web documents in English language that were crawled in 2009
  (CW).

\end{itemize}

These two are extremes: NYT is a well-curated, relatively clean,
longitudinal corpus, i.e., documents therein have a clear structure,
use proper language with few typos, and cover a long time period. CW
is a ``World Wild Web'' corpus, i.e., documents therein are highly
heterogeneous in structure, content, and language.

For NYT a document consists of the newspaper article's title and
body. To make CW more handleable, we use boilerplate detection as
described by Kohlsch\"utter et al.~\cite{Kohlschutter:2010fk} and
implemented in boilerpipe's~\cite{boilerpipe} default extractor, to
identify the core content of documents. On both datasets, we use
OpenNLP~\cite{opennlp} to detect sentence boundaries in
documents. Sentence boundaries act as barriers, i.e., we do not
consider $n$-grams that span across sentences in our experiments. As
described in Section~\ref{sec:impl--extens}, in a pre-processing step,
we convert both datasets into sequences of integer
term-identi\-fiers. The term dictionary is kept as a single text file;
documents are spread as key-value pairs of 64-bit document identifier
and content integer array over a total of 256 binary
files. Table~\ref{tab:Datasets} summarizes characteristics of the two
datasets.

\vspace{-1mm}
\subsection{Output Characteristics}
\vspace{-1mm}

\begin{figure*}[t]
  \centering
  \textbf{NYT}\hspace{50mm}\textbf{CW}\\
  \vspace{-2mm}
  \includegraphics[width=0.30\textwidth,trim=5mm 5mm 5mm 5mm]{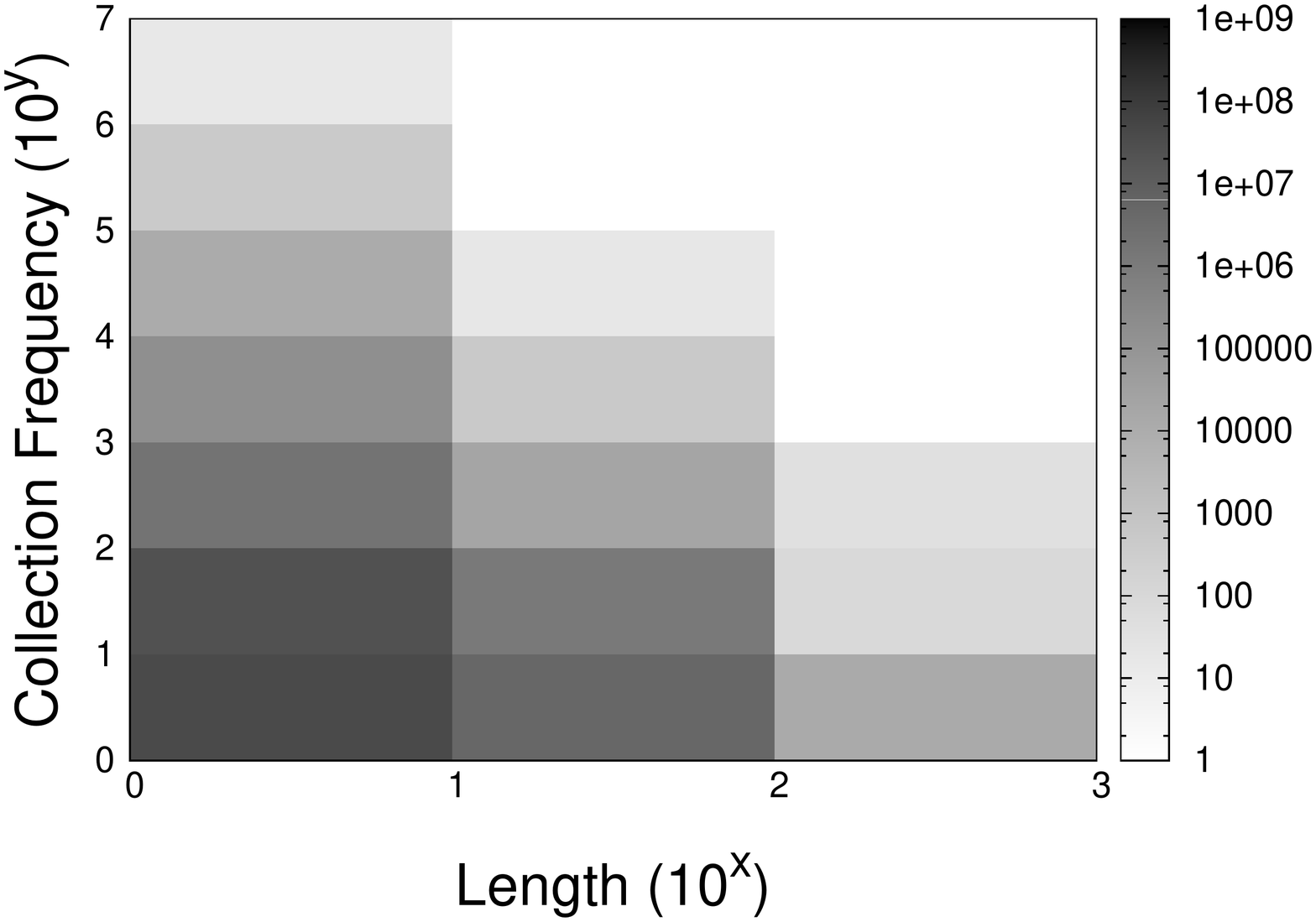}
  \includegraphics[width=0.30\textwidth,trim=5mm 5mm 5mm 5mm]{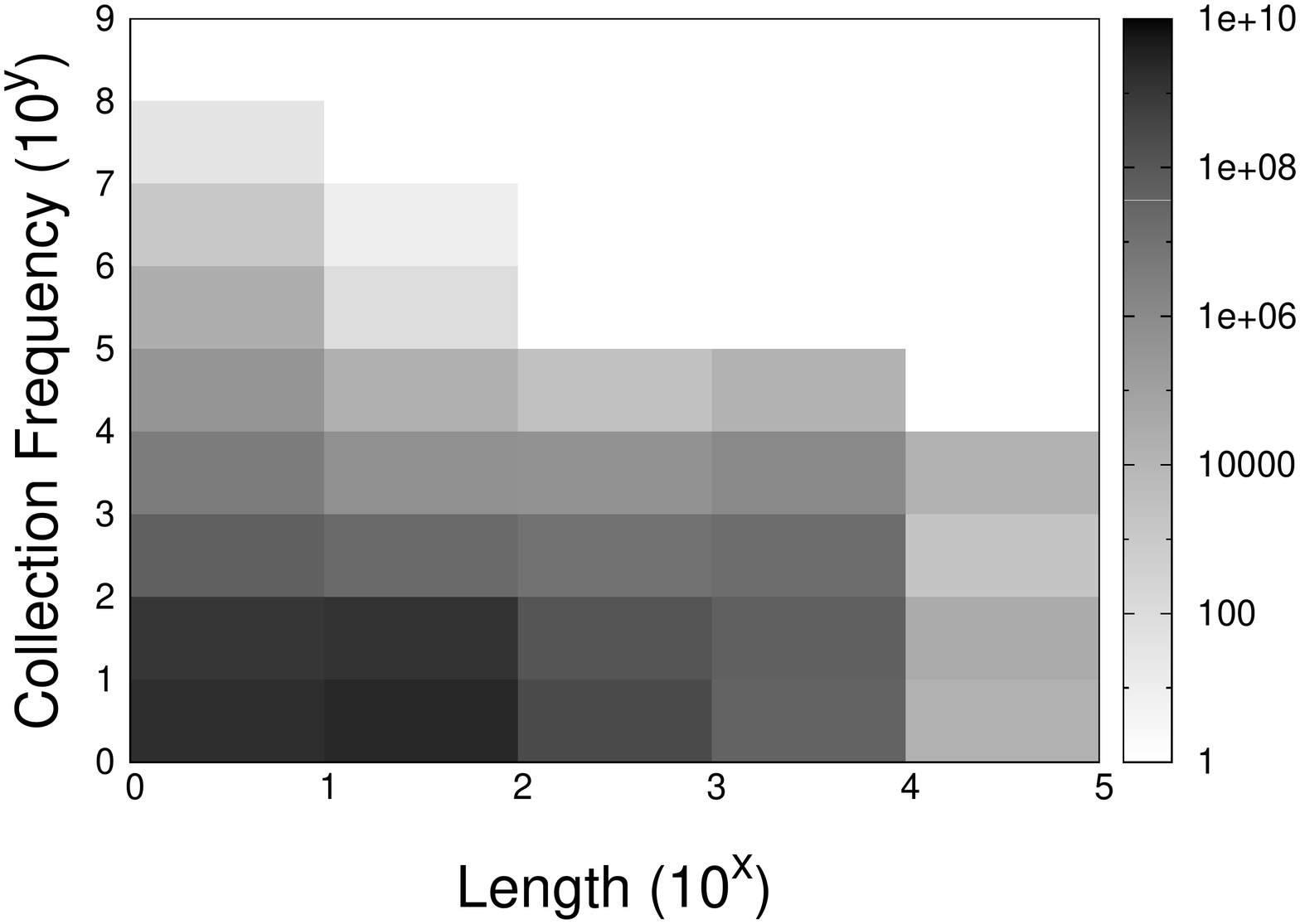}
  \vspace{-2mm}
  \caption{Output characteristics as \# of $n$-grams $\textbf{s}$ with
    $cf(\mathbf{s}) \ge 5$ per $n$-gram length and collection
    frequency}
  \label{fig:Output}
\end{figure*}

Let us first look at the $n$-gram statistics that (or, parts of which)
we expect as output from all methods. To this end, for both document
collections, we determine all $n$-grams that occur at least five times
(i.e., $\tau=5$ and $\sigma=\infty$). We bin $n$-grams into
2-dimensional buckets of exponential width, i.e., the $n$-gram
$\mathbf{s}$ with collection frequency $cf(\mathbf{s})$ goes into
bucket $(i,j)$ where $i= \lfloor log_{10}\,|\mathbf{s}|\rfloor$ and
$j=\lfloor log_{10}\,cf(\mathbf{s}) \rfloor$. Figure~\ref{fig:Output}
reports the number of $n$-grams per bucket.

The figure reveals that the distribution is biased toward short and
less frequent $n$-grams. Consequently, as we lower the value of
$\tau$, all methods have to deal with a drastically increasing number
of $n$-grams. What can also be seen from Figure~\ref{fig:Output} is
that, in both datasets, $n$-grams exist that are very long, containing
hundred or more terms, and occur more than ten times in the document
collection. Examples of long $n$-grams that we see in the output
include ingredient lists of recipes (e.g.,\smalltt{\ldots 1 tablespoon
  cooking oil\ldots}) and chess openings (e.g., \smalltt{e4 e5 2
  nf3\ldots}) in NYT; in CW they include web spam (e.g.,
\smalltt{travel tips san miguel tourism san miguel transport san
  miguel\ldots}) as well as error messages and stack traces from web
servers and other software (e.g., \smalltt{\ldots php on line 91
  warning\ldots}) that also occur within user discussions in
forums. For the \textsc{Apriori}-based methods, such long $n$-grams
are unfavorable, since they require many iterations to identify them.

\vspace{-1mm}
\subsection{Use Cases}
\vspace{-1mm}

As a first experiment, we investigate how the methods perform for
parameter settings chosen to reflect two typical use cases, namely,
\textit{training a language model} and \textit{text analytics}. For
the first use case, we set $\tau=10$ on NYT and $\tau=100$ on CW, as
relatively low minimum collection frequencies, in combination with
$\sigma=5$. The $n$-gram statistics made public by
Google~\cite{googlengrams}, as a comparison, were computed with
parameter settings $\tau=40$ and $\sigma=5$ on parts of the Web. For
the second use case, we choose $\sigma=100$, as a relatively high
maximum sequence length, combined with $\tau=100$ on NYT and
$\tau=1,000$ on CW. The idea in the analytics use case is to identify
recurring fragments of text (e.g., quotations or idioms) to be
analyzed further (e.g., their spread over time).

\begin{figure*}[t]
  \centering
  \vspace{-2mm}
  \subfigure[]{\label{fig:UCLM}\includegraphics[width=0.30\textwidth,trim=5mm 5mm 5mm 5mm]{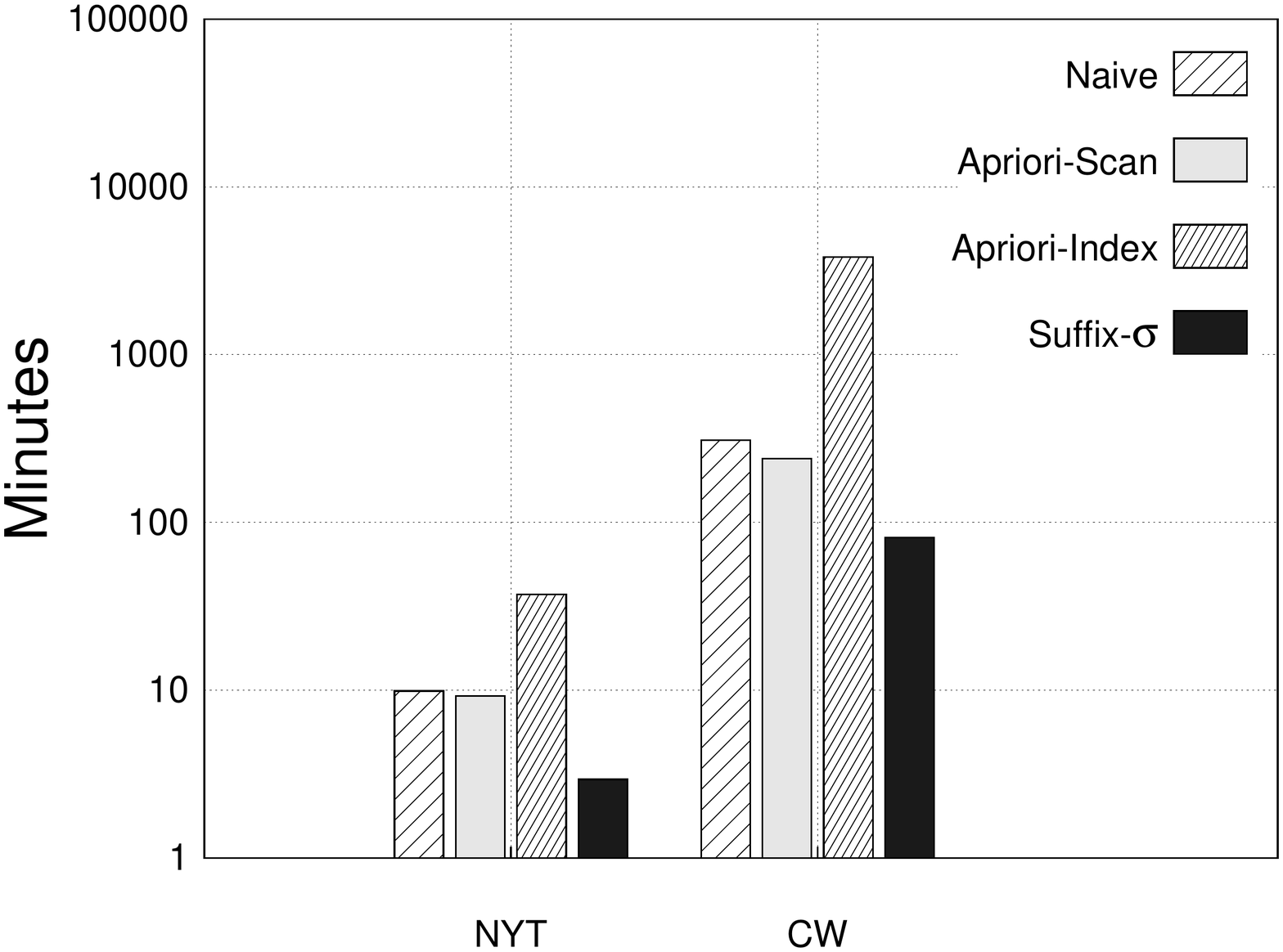}}
  \subfigure[]{\label{fig:UCSMA}\includegraphics[width=0.30\textwidth,trim=5mm 5mm 5mm 5mm]{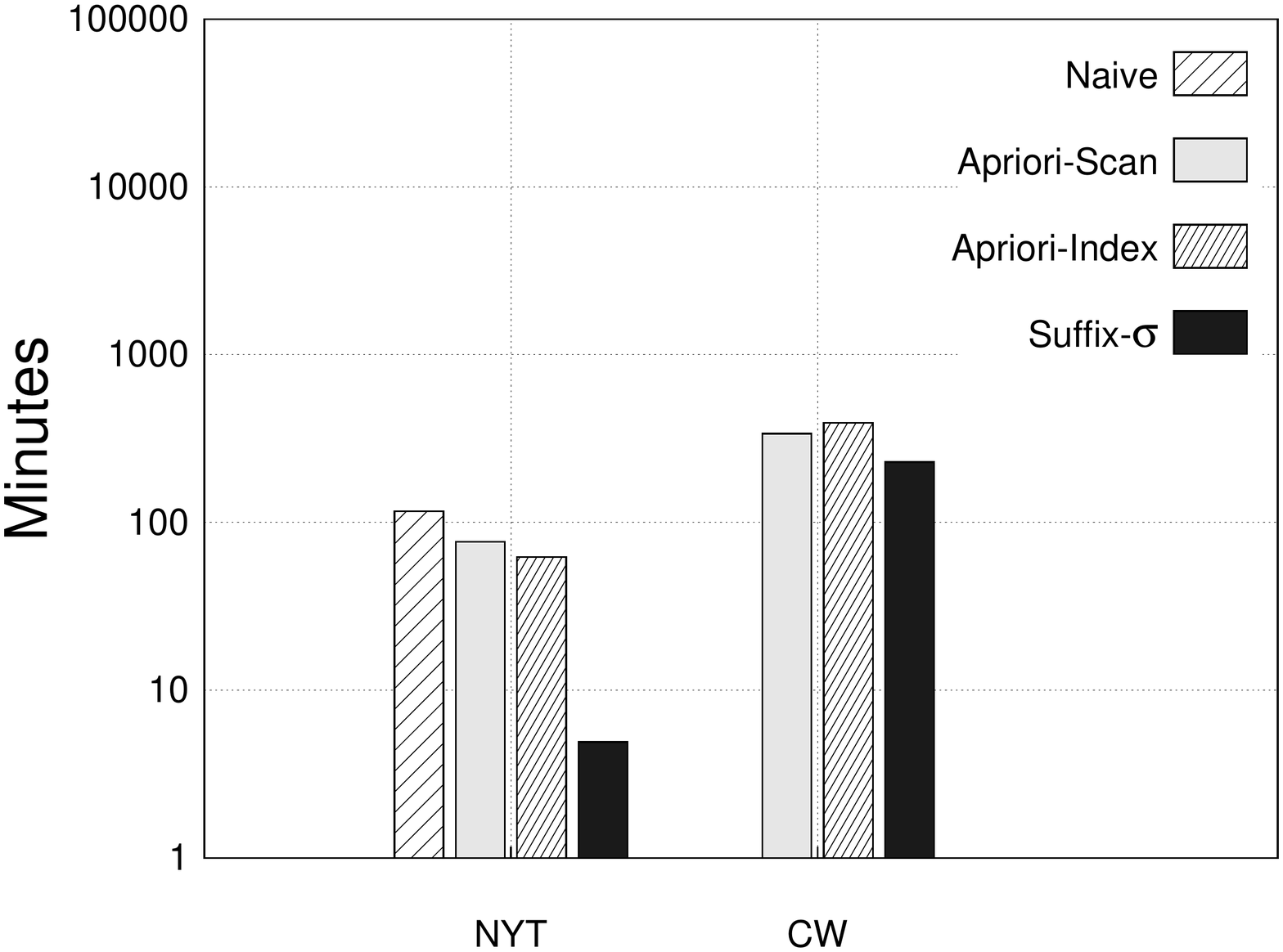}}
  \vspace{-2mm}
  \caption{Wallclock times in minutes for (a) \textit{training a
      language model} ($\sigma=5$, NYT: $\tau=10$ / CW: $\tau=100$)
    and (b) \textit{text analytics} ($\sigma=100$, NYT: $\tau=100$ /
    CW: $\tau=1,000$) as two typical use cases}
  \vspace{-2mm}
  \label{fig:UseCases}
\end{figure*}

Figure~\ref{fig:UseCases} reports wallclock-time measurements obtained
for these two use cases with 64~map/reduce slots. For our
language-model use case, \textsc{Suffix}-$\sigma$ outperforms
\textsc{Apriori-Scan} as the best competitor by a \textit{factor 3x}
on both datasets. For our analytics use case, we see a \textit{factor
  12x} improvement over \textsc{Apriori-Index} as the best competitor
on NYT; on CW \textsc{Suffix}-$\sigma$ still outperforms the next best
\textsc{Apriori-Scan} by a \textit{factor~1.5x}. Measurements for
\textsc{Na\"ive} on CW in are missing, since the method did not
complete in reasonable time.

\vspace{-1mm}
\subsection{Varying Minimum Collection Frequency}
\vspace{-1mm}

\begin{figure*}[t]
  \centering
  \textbf{NYT}\\
  \vspace{-3mm} \subfigure[Wallclock times]{\label{fig:MSNYTWC}\includegraphics[width=0.30\textwidth,trim=5mm 5mm 5mm 5mm]{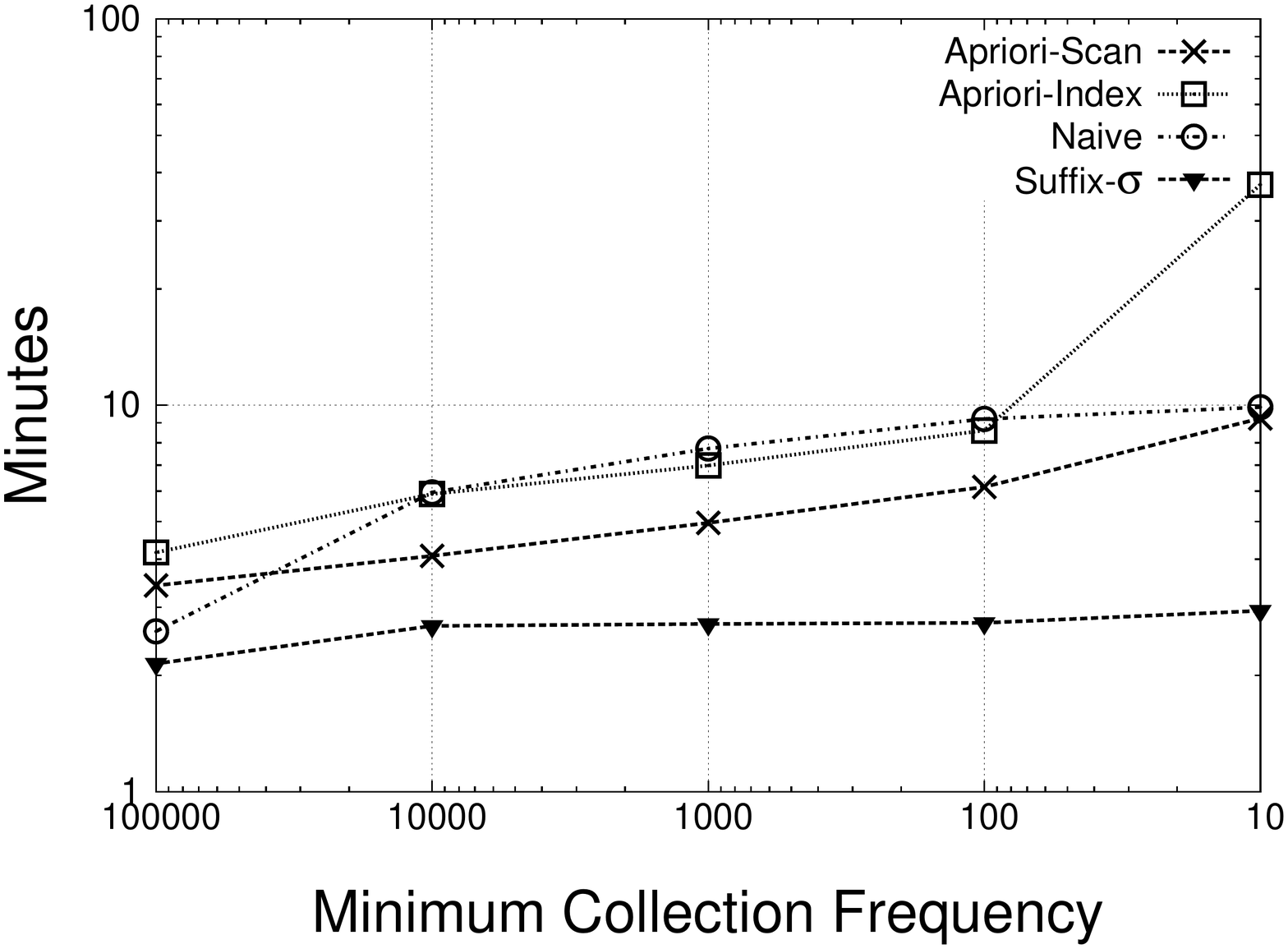}}
  \subfigure[Bytes transferred]{\label{fig:MSNYTBytes}\includegraphics[width=0.30\textwidth,trim=5mm 5mm 5mm 5mm]{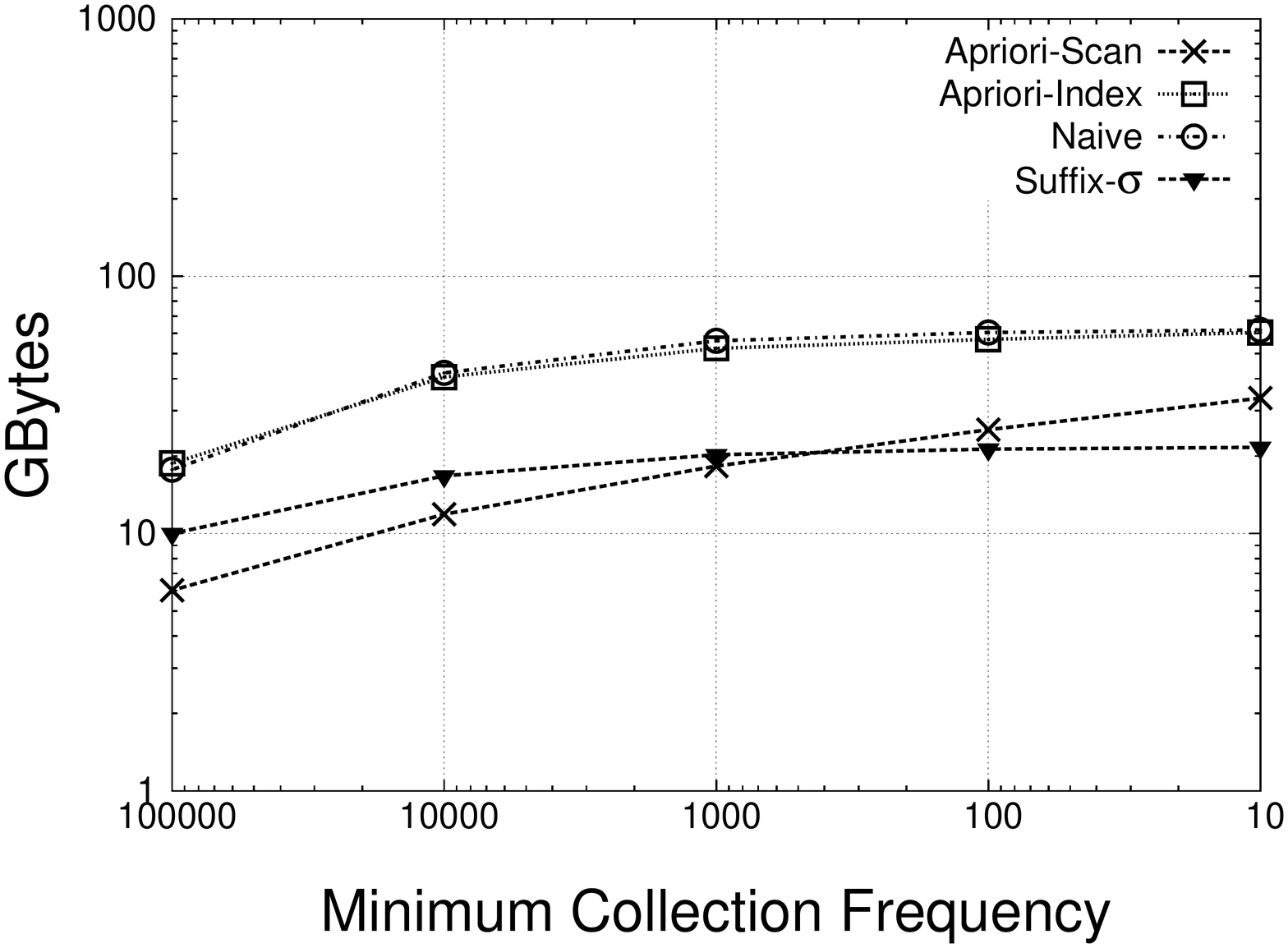}}
  \subfigure[\# of records]{\label{fig:MSNYTRecords}\includegraphics[width=0.30\textwidth,trim=5mm 5mm 5mm 5mm]{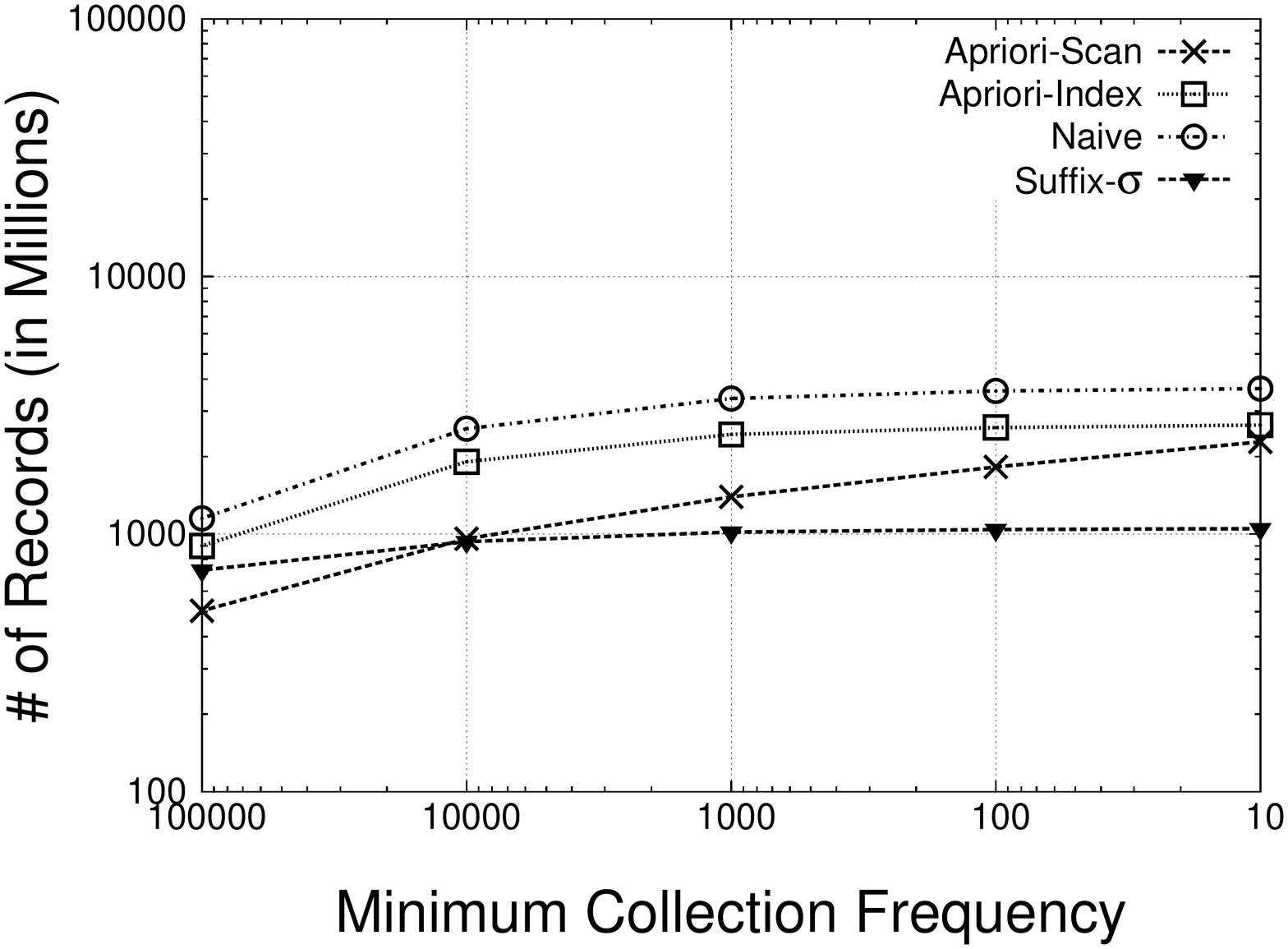}}

  \vspace{2mm}

  \textbf{CW}\\
  \vspace{-3mm} \subfigure[Wallclock times]{\label{fig:MSCWWC}\includegraphics[width=0.30\textwidth,trim=5mm 5mm 5mm 5mm]{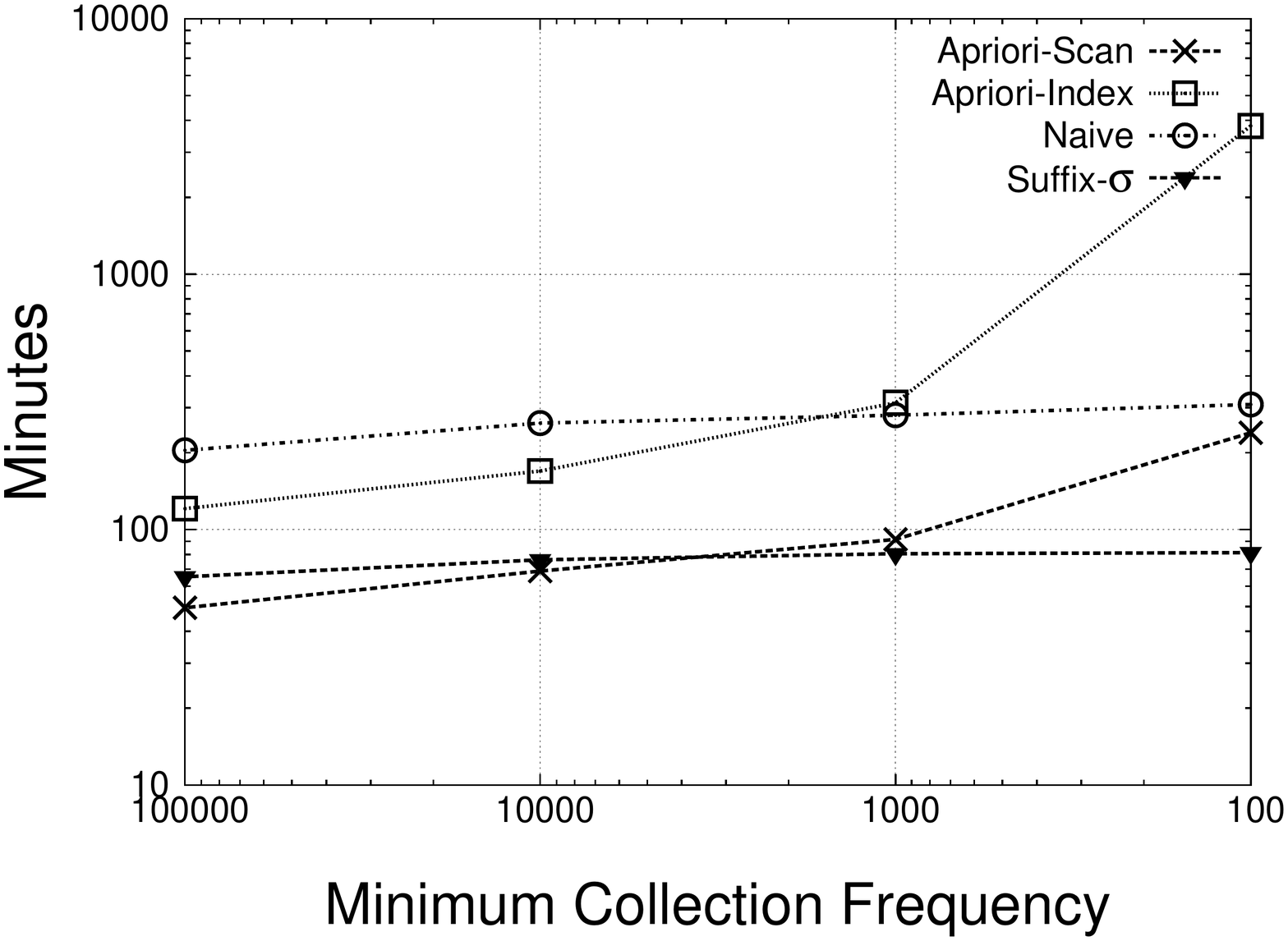}}
  \subfigure[Bytes transferred]{\label{fig:MSCWBytes}\includegraphics[width=0.30\textwidth,trim=5mm 5mm 5mm 5mm]{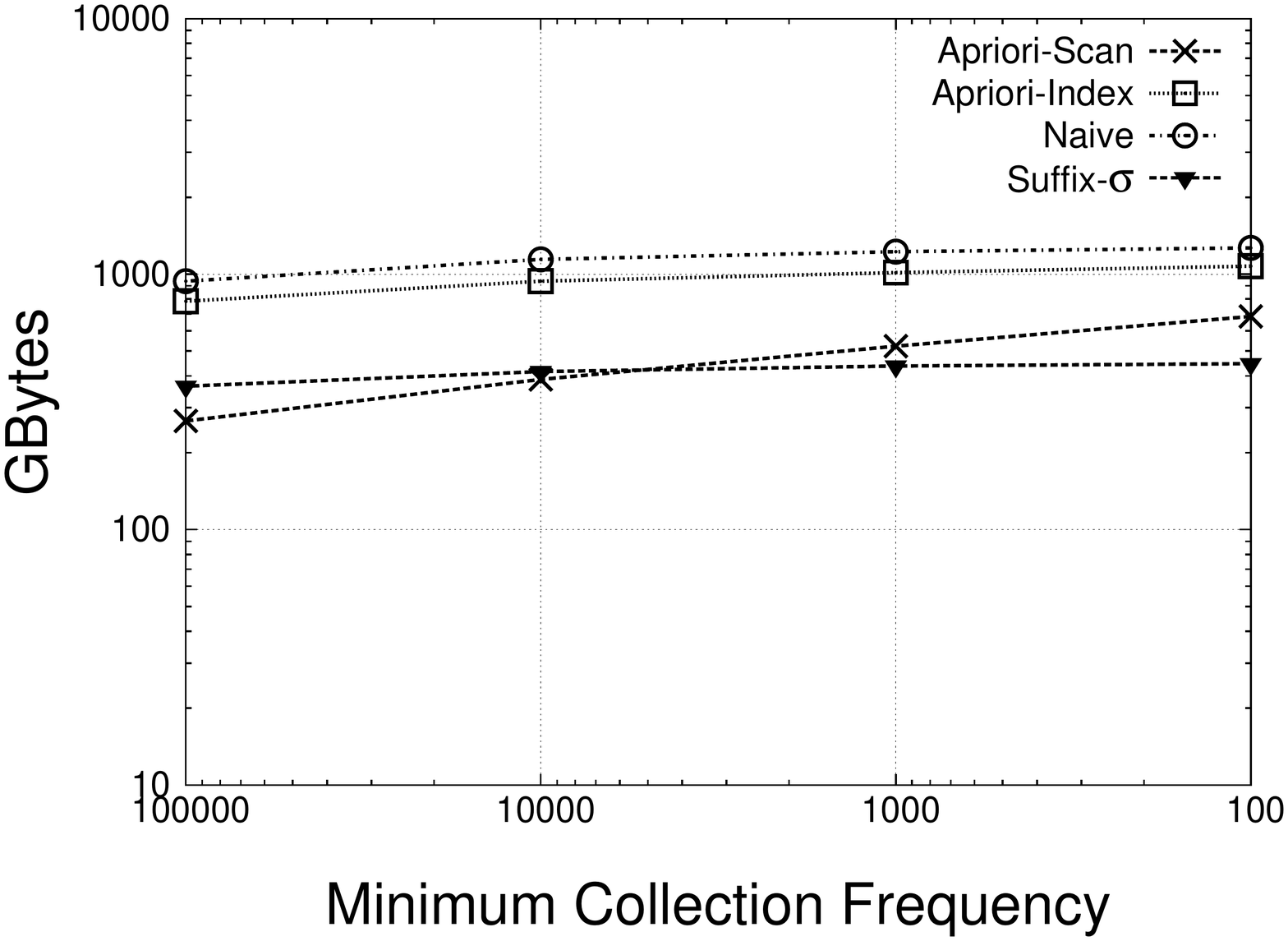}}
  \subfigure[\# of records]{\label{fig:MSCWRecords}\includegraphics[width=0.30\textwidth,trim=5mm 5mm 5mm 5mm]{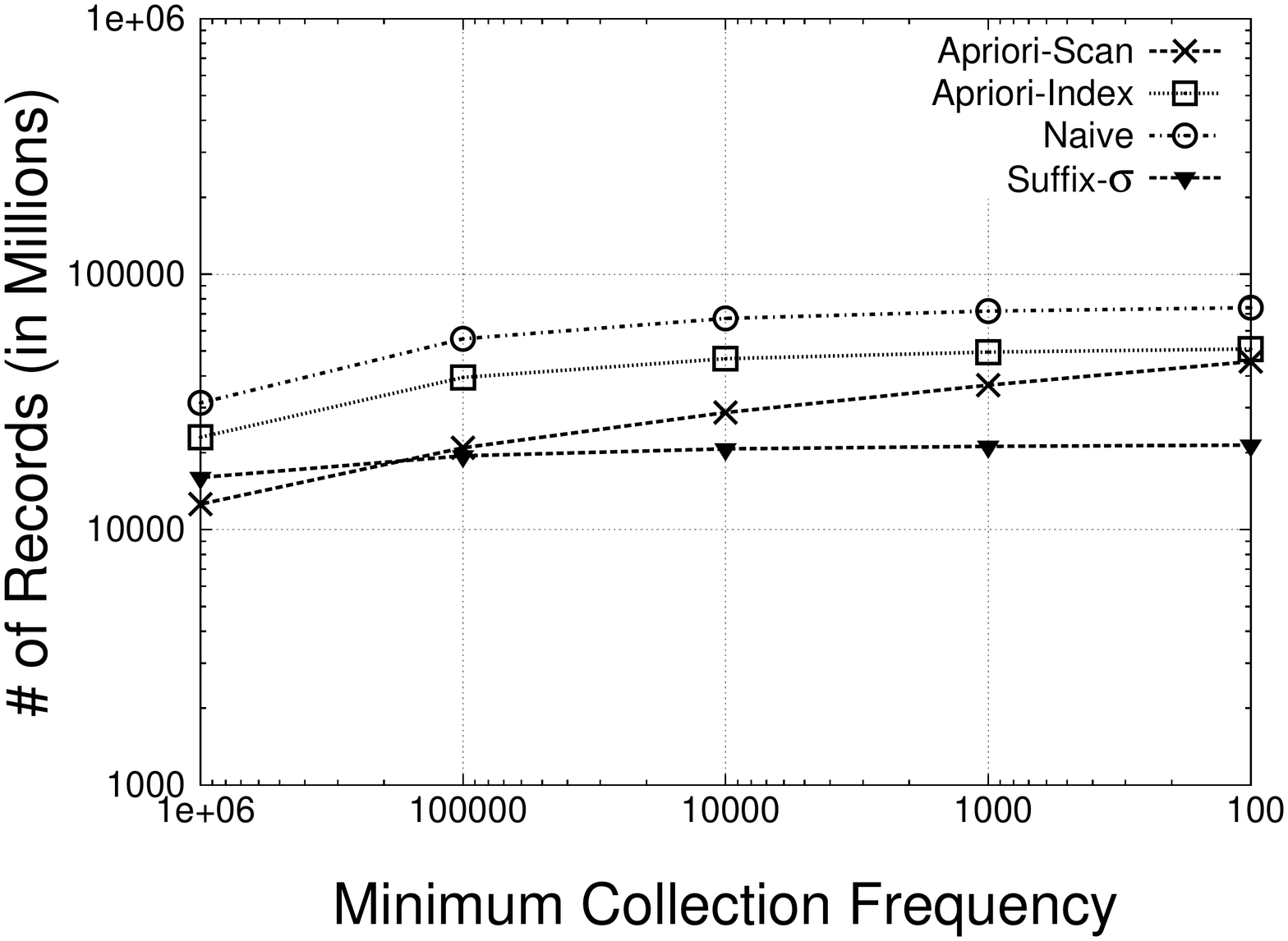}}
  \vspace{-2mm}
  \caption{Varying the minimum collection frequency $\tau$}
  \label{fig:VaryingMinimumSupport}
  \vspace{-2mm}
\end{figure*}

Our second experiment studies how the methods behave as we vary the
minimum collection frequency $\tau$. We use a maximum length
$\sigma=5$ and apply all methods to the entire datasets. Measurements
are performed using 64~map/reduce slots and reported in
Figure~\ref{fig:VaryingMinimumSupport}.

We observe that for high minimum collection frequencies,
\textsc{Suffix}-$\sigma$ performs as well as the best competitor
\textsc{Apriori-Scan}. For low minimum collection frequencies, it
significantly outperforms the other methods. Both
\textsc{Apriori}-based method show steep increases in wallclock time
as we lower the minimum collection frequency -- especially when we
reach the lowest value of $\tau$ on each document collection. This is
natural, because for both methods the work that has to be done in the
$k$-th iteration depends on the number of $(k-1)$-grams output in the
previous iteration, which have to be joined or kept in a dictionary,
as described in Section~\ref{sec:methods-based-prior}. As observed in
Figure~\ref{fig:Output} above, the number of $k$-grams grows
drastically as we decrease the value of $\tau$. When looking at the
number of bytes and the number of records transferred, we see
analogous behavior. For low values of $\tau$, \textsc{Suffix}-$\sigma$
transfers significantly less data than its competitors.

\vspace{-1mm}
\subsection{Varying Maximum Length}
\vspace{-1mm}

\begin{figure*}[t]
  \centering
  \textbf{NYT}\\
  \vspace{-3mm}
  \subfigure[Wallclock times]{\label{fig:MLNYTWC}\includegraphics[width=0.30\textwidth,trim=5mm 5mm 5mm 5mm]{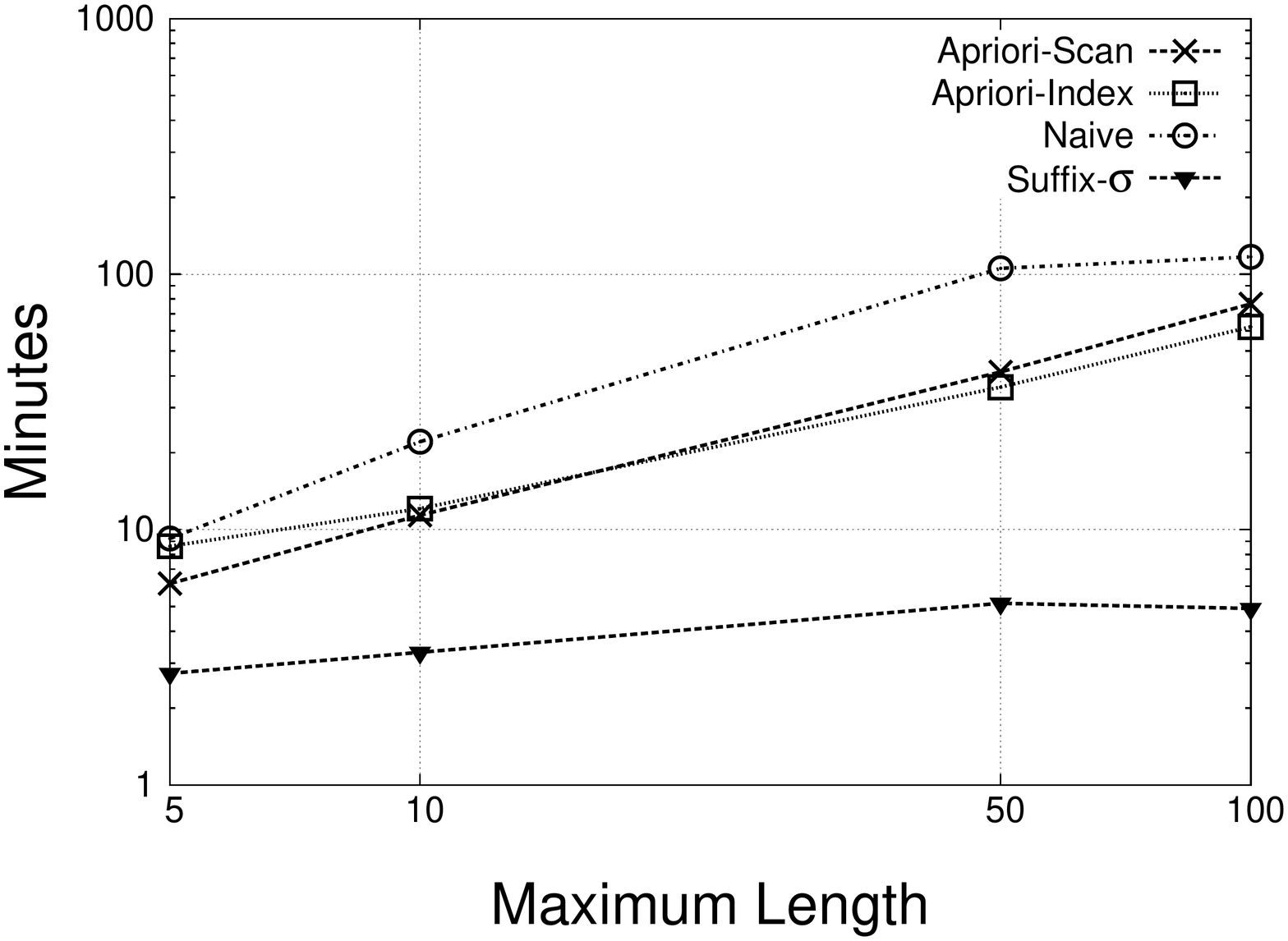}}
  \subfigure[Bytes transferred]{\label{fig:MLNYTBytes}\includegraphics[width=0.30\textwidth,trim=5mm 5mm 5mm 5mm]{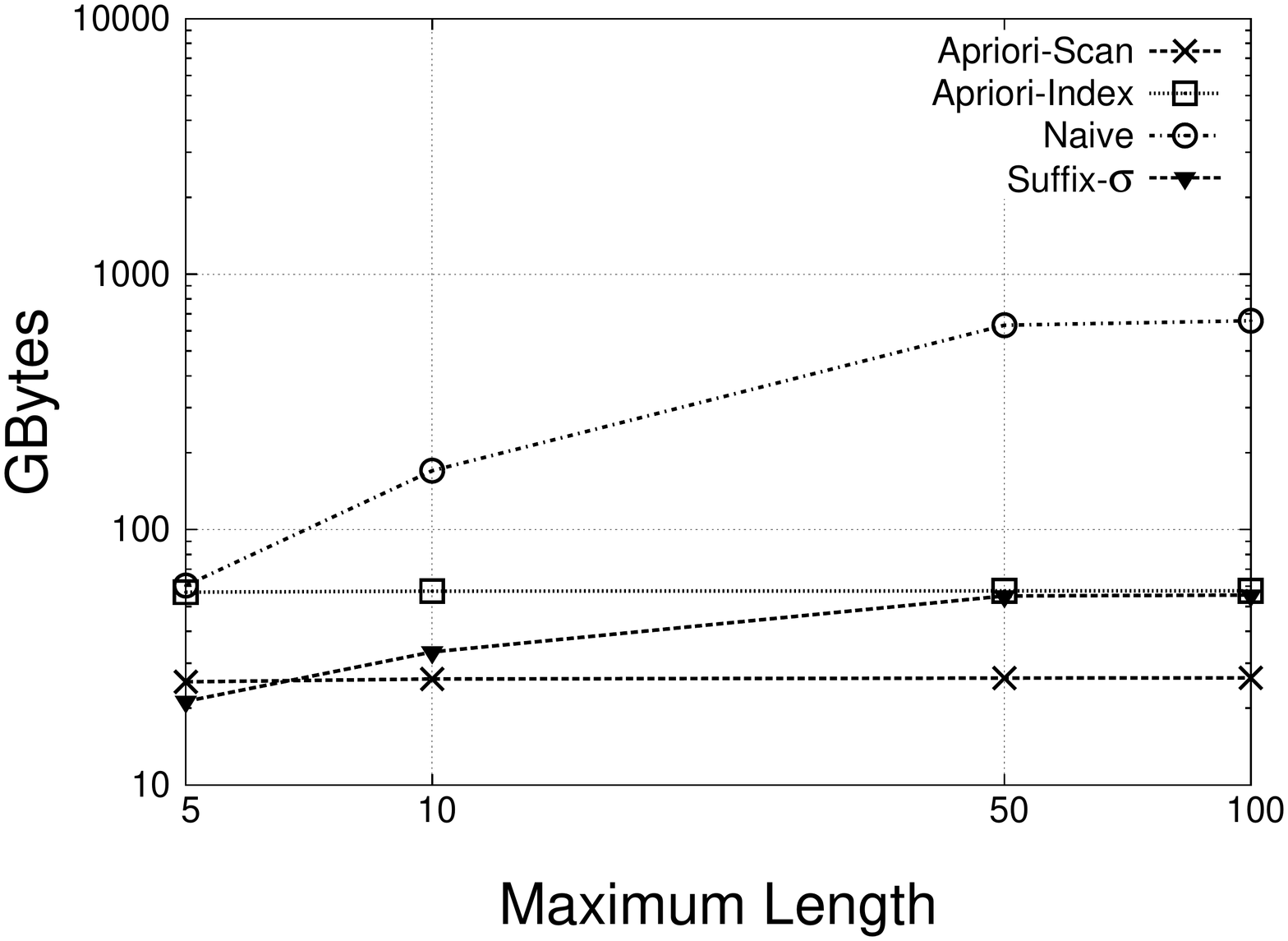}}
  \subfigure[\# of records]{\label{fig:MLNYTRecords}\includegraphics[width=0.30\textwidth,trim=5mm 5mm 5mm 5mm]{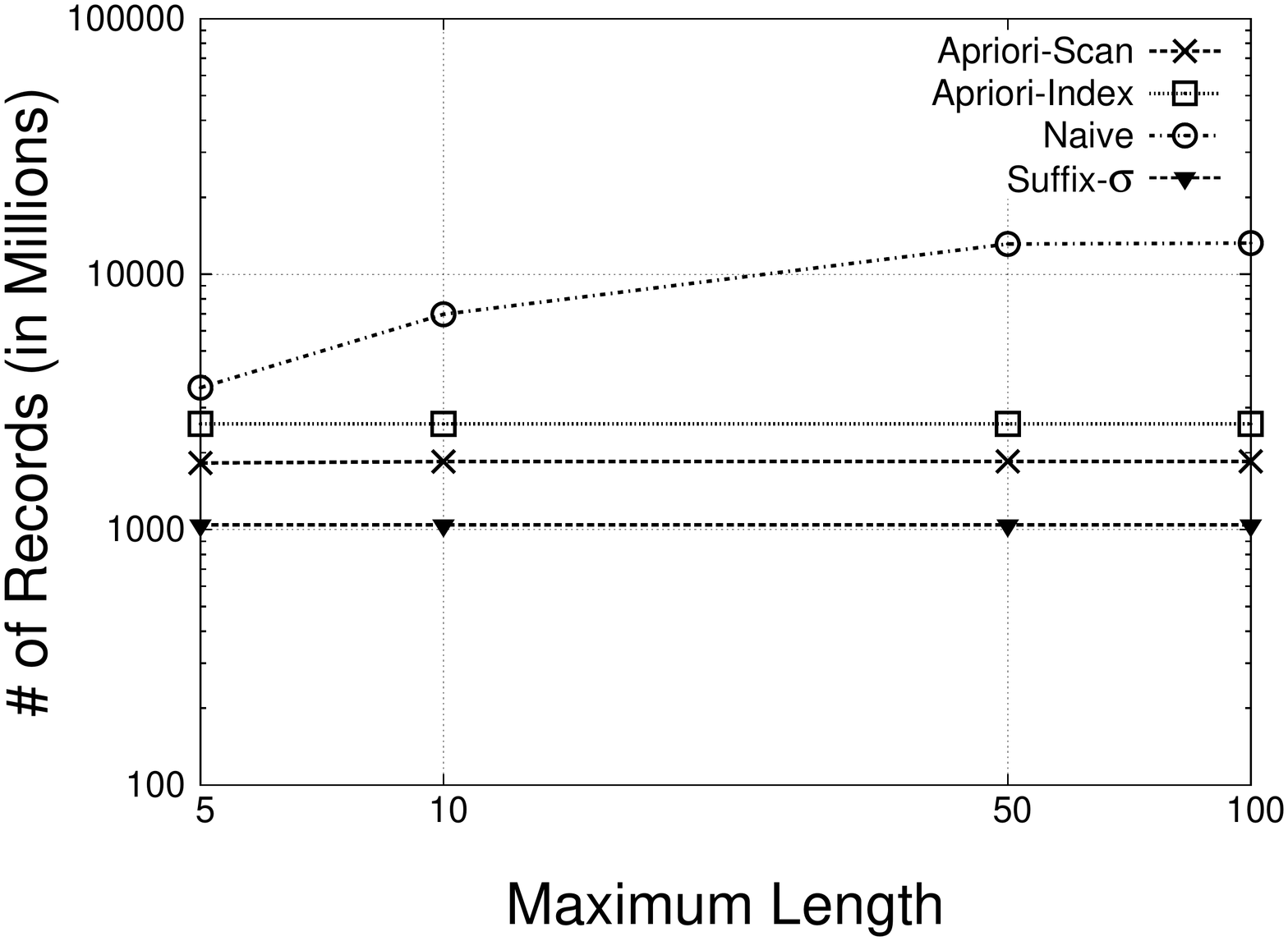}}

  \vspace{2mm}

  \textbf{CW}\\
  \vspace{-3mm}
  \subfigure[Wallclock times]{\label{fig:MLNYTWC}\includegraphics[width=0.30\textwidth,trim=5mm 5mm 5mm 5mm]{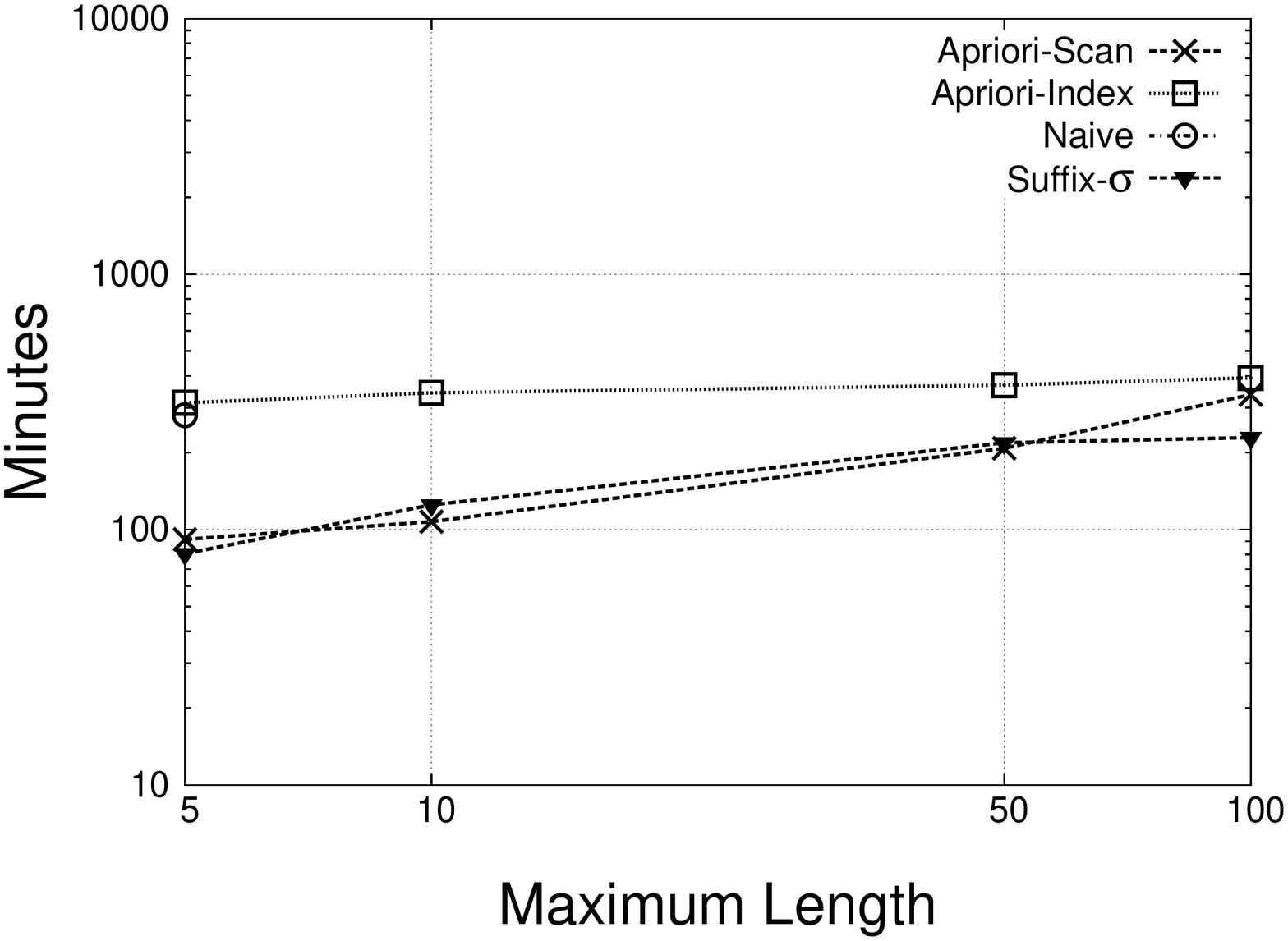}}
  \subfigure[Bytes transferred]{\label{fig:MLNYTBytes}\includegraphics[width=0.30\textwidth,trim=5mm 5mm 5mm 5mm]{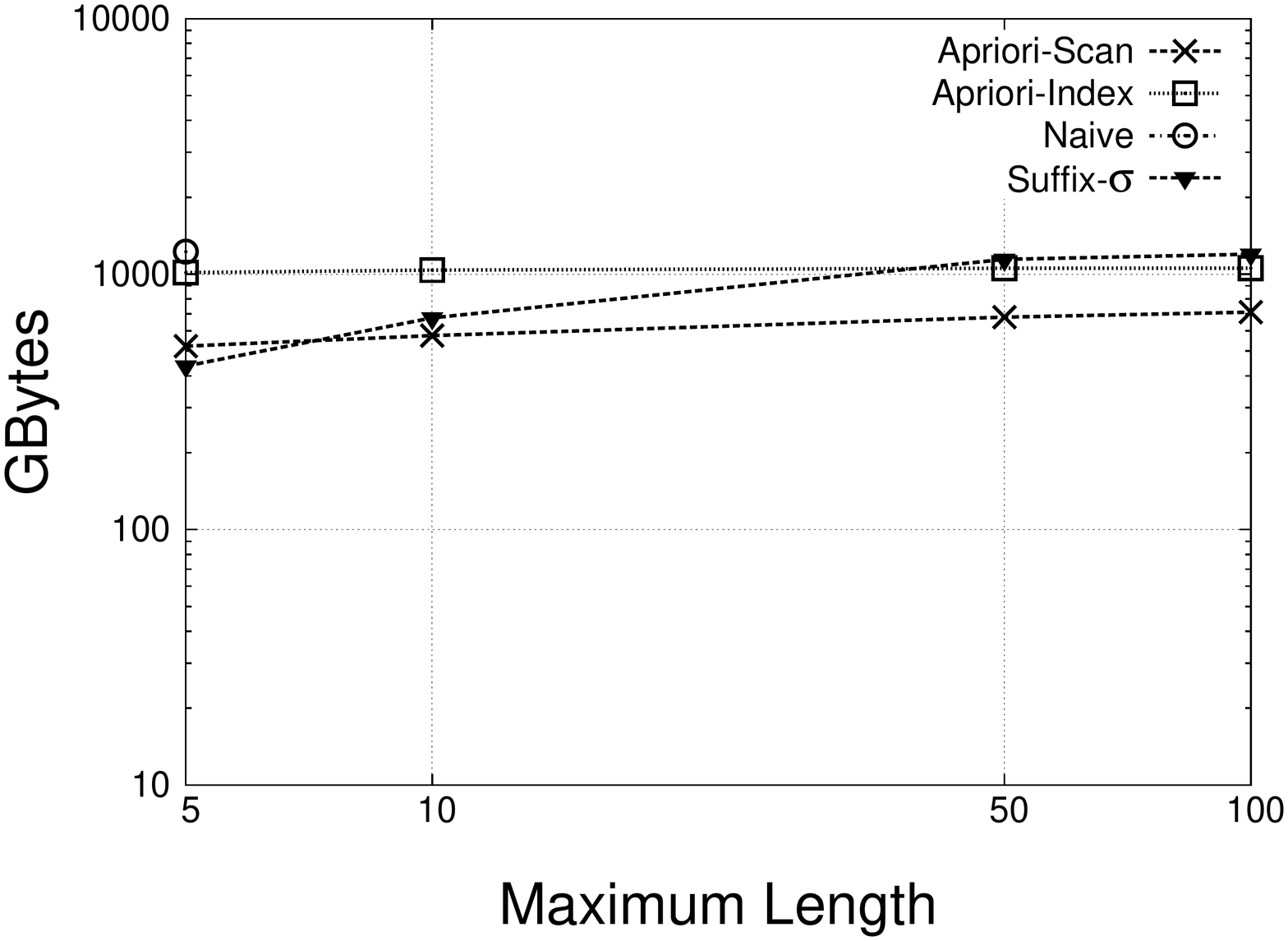}}
  \subfigure[\# of records]{\label{fig:MLNYTRecords}\includegraphics[width=0.30\textwidth,trim=5mm 5mm 5mm 5mm]{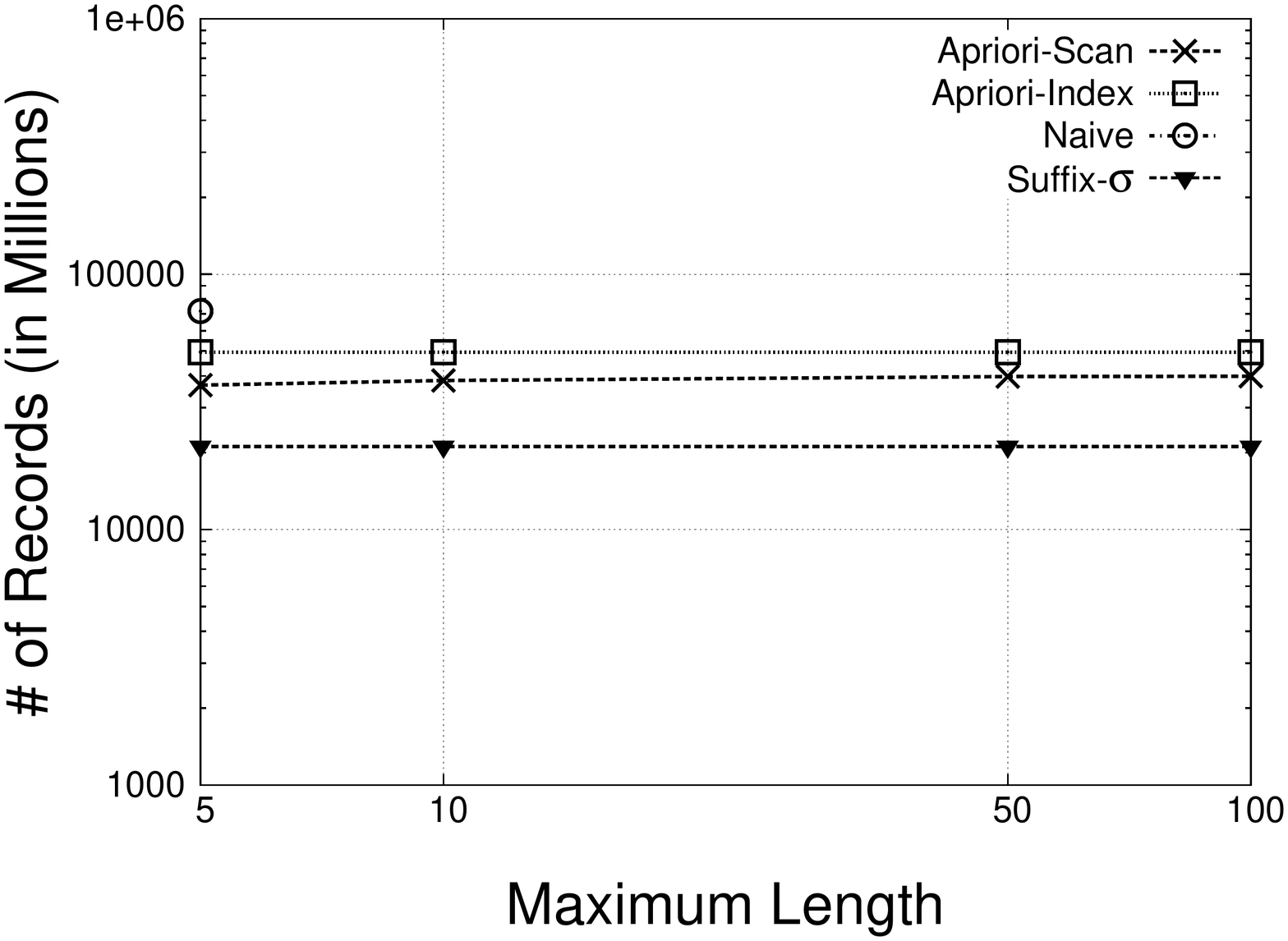}}
  \vspace{-2mm}
  \caption{Varying the maximum length $\sigma$}
  \label{fig:VaryingMaximumLength}
  \vspace{-2mm}
\end{figure*}

In this third experiment, we study the methods' behavior as we vary
the maximum length $\sigma$. The minimum collection frequency is set
as $\tau=100$ for NYT and $\tau=1,000$ for CW to reflect their
different scale. Measurements are performed on the entire datasets
with 64~map/reduce slots and reported in
Figure~\ref{fig:VaryingMaximumLength}. Measurements for $\sigma > 5$
are missing for \textsc{Na\"ive} on CW, since the method did not
finish within reasonable time for those parameter settings.

\textsc{Suffix}-$\sigma$ is on par with the best-performing competitor
on CW, when considering $n$-grams of length up to $50$. For $\sigma =
100$, it outperforms the next best \textsc{Apriori-Scan} by a
\textit{factor 1.5x}. On NYT, Suffix-$\sigma$ consistently outperforms
all competitors by a wide margin. When we increase the value of
$\sigma$, the \textsc{Apriori}-based methods need to run more Hadoop
jobs, so that their wallclock times keep increasing. For
\textsc{Na\"ive} and \textsc{Suffix}-$\sigma$, on the other hand, we
observe a saturation of wallclock times. This is expected, since these
methods have to do additional work only for input sequences longer
than $\sigma$ consisting of terms that occur at least $\tau$ times in
the document collection. When looking at the number of bytes and the
number of records transferred, we observe a saturation for
\textsc{Na\"ive} for the reason mentioned above. For
\textsc{Suffix}-$\sigma$ only the number of bytes saturates, the
number of records transferred is constant, since it depends only on
the minimum collection frequency $\tau$. Further, we see that
\textsc{Suffix}-$\sigma$ consistently transfers fewest records.

\vspace{-1mm}
\subsection{Scaling the Datasets}
\vspace{-1mm}

Next, we investigate how the methods react to changes in the scale of
the datasets. To this end, both from NYT and CW, we extract smaller
datasets that contain a random $25\%$, $50\%$, or $75\%$ subset of the
documents. Again, the minimum collection frequency is set as
$\tau=100$ for NYT and $\tau=1,000$ for CW. The maximum length is set
as $\sigma=5$. Wallclock times are measured using 64~map/reduce slots.

\begin{figure*}[t]
  \centering
  \textbf{NYT}\hspace{50mm}\textbf{CW}\\
  \vspace{-3mm}
  \subfigure[Wallclock times]{\label{fig:DSNYTWC}\includegraphics[width=0.30\textwidth,trim=5mm 5mm 5mm 5mm]{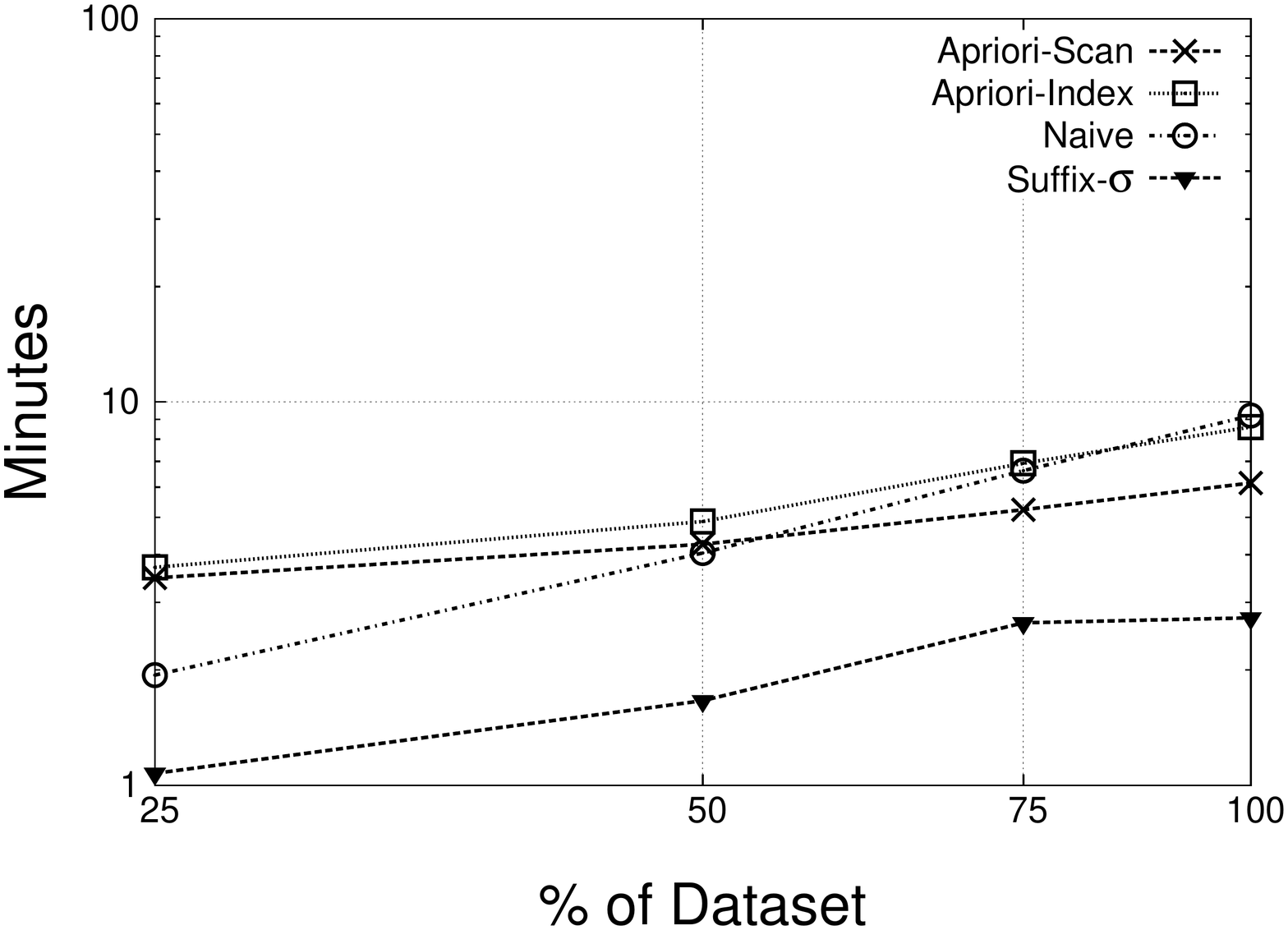}}
  \subfigure[Wallclock times]{\label{fig:DSCWWC}\includegraphics[width=0.30\textwidth,trim=5mm 5mm 5mm 5mm]{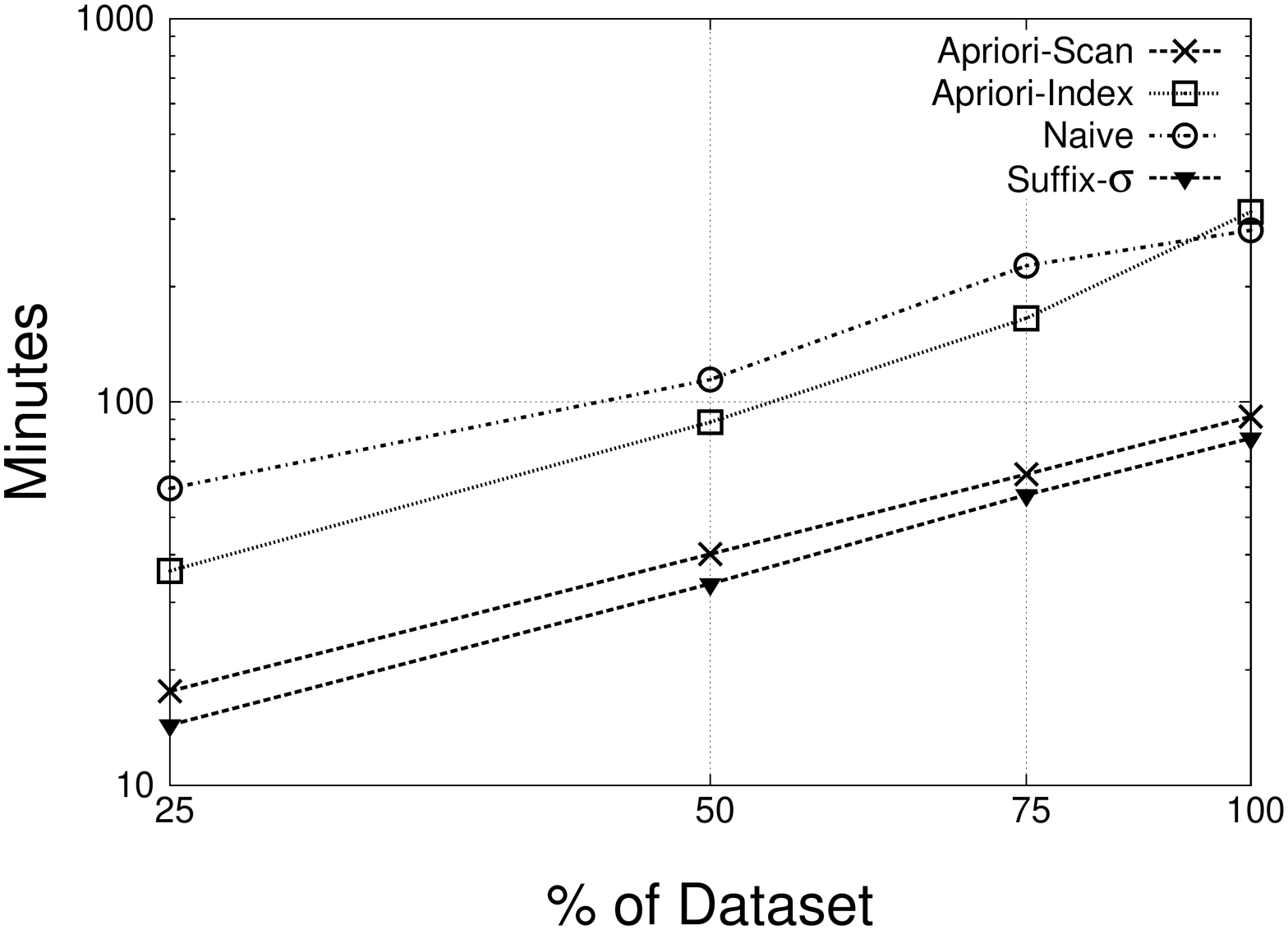}}
  \vspace{-2mm}
  \caption{Scaling the datasets}
  \label{fig:ScalingDataset}
  \vspace{-2mm}
\end{figure*}

From Figure~\ref{fig:ScalingDataset}, we observe that \textsc{Na\"ive}
handles additional data equally well on both datasets. The other
methods' scalability is comparable to that of \textsc{Na\"ive} on CW,
as can be seen from their almost-identical slopes. On NYT, in
contrast, \textsc{Apriori-Scan}, \textsc{Apriori-Index}, and
\textsc{Suffix}-$\sigma$ cope slightly better with additional data
than \textsc{Na\"ive}. This is due to the different characteristics of
the two datasets.




\vspace{-1mm}
\subsection{Scaling Computational Resources}
\vspace{-1mm}

Our final experiment explores how the methods behave as we scale
computational resources. Again, we set $\tau=100$ for NYT and
$\tau=1,000$ for CW. All methods are applied to the $50\%$ samples of
documents from the collections. We vary the number of map/reduce slots
as 16, 32, 48, and 64. The number of cluster nodes remains constant in
this experiment, since we cannot add/remove machines to/from the
cluster due to organizational restrictions. We thus only vary the
amount of parallel work every machine can do; their total number
remains constant throughout this experiment.

\begin{figure*}[t]
  \centering
  \textbf{NYT}\hspace{50mm}\textbf{CW}\\
  \vspace{-3mm}
  \subfigure[Wallclock times]{\label{fig:DCNYTWC}\includegraphics[width=0.30\textwidth,trim=5mm 5mm 5mm 5mm]{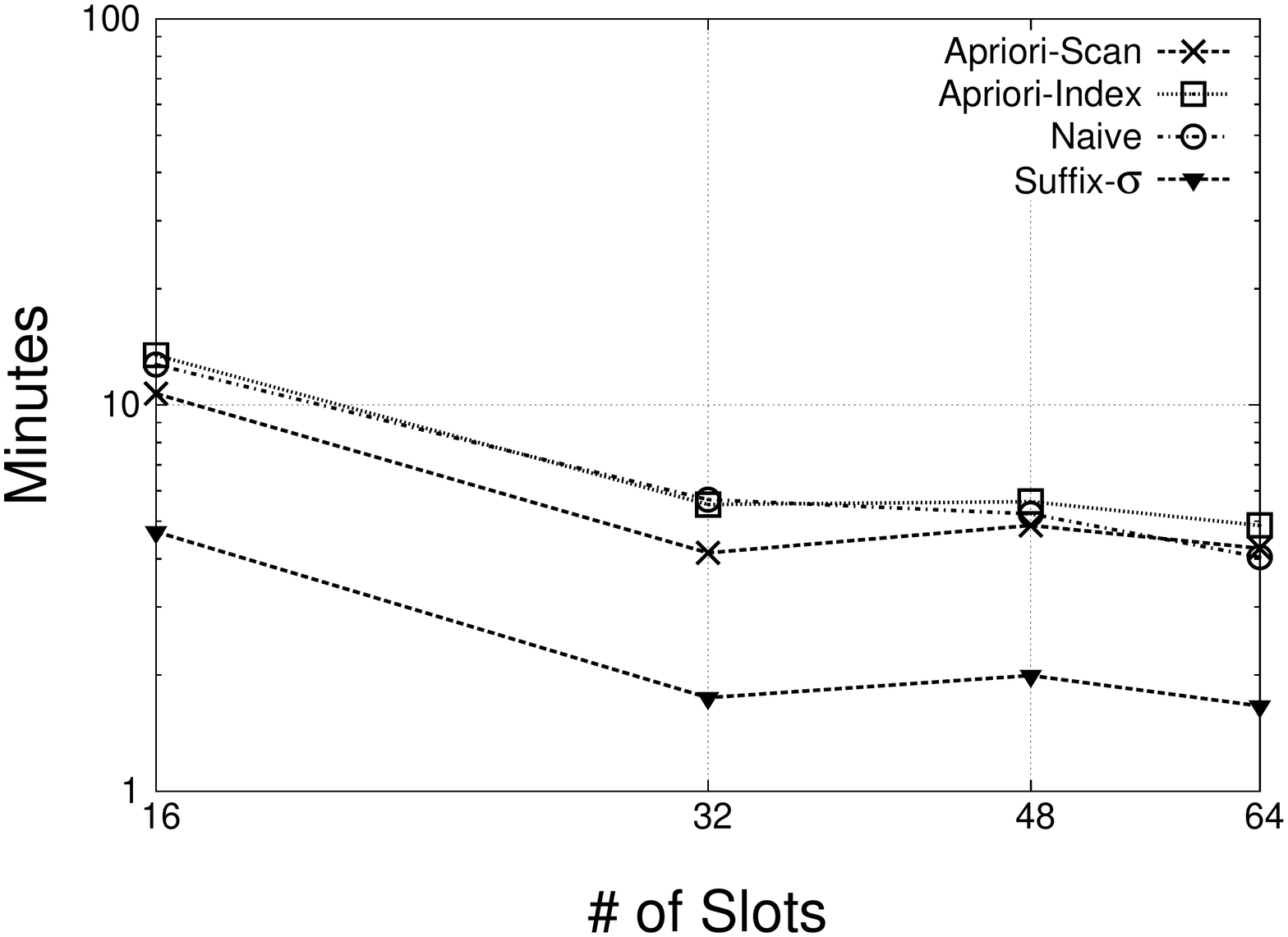}}
  \subfigure[Wallclock times]{\label{fig:DCCWWC}\includegraphics[width=0.30\textwidth,trim=5mm 5mm 5mm 5mm]{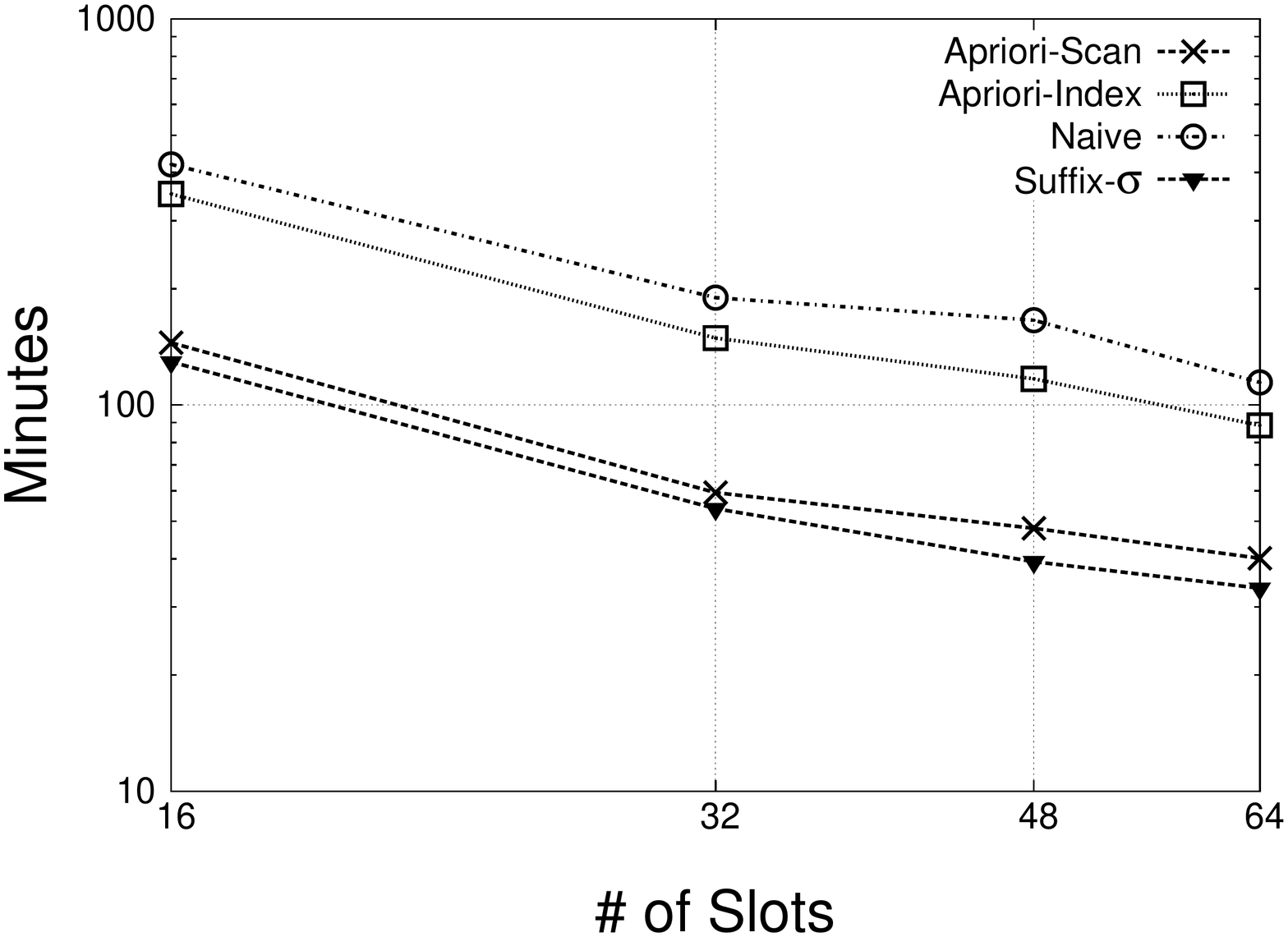}}
  \vspace{-2mm}
  \caption{Scaling computational resources}
  \label{fig:ScalingCluster}
  \vspace{-2mm}
\end{figure*}

We observe from Figure~\ref{fig:ScalingCluster} that all methods show
comparable behavior as we make additional computational resources
available. Or, put differently, all methods make equally effective use
of them. What can also be observed across all methods is that the
gains of adding more computational resources are diminishing --
because of mappers and reducers competing for shared devices such as
hard disks and network interfaces. This phenomenon is more pronounced
on NYT than CW, since methods take generally less time on the smaller
dataset, so that competition for shared devices is fiercer and has no
chance to level out over time.

\vspace{-2mm}
\subsection*{Summary}
\vspace{-1mm}

What we see in our experiments is that \textsc{Suffix}-$\sigma$
outperforms its competitors when long and/or less frequent $n$-grams
are considered. Even otherwise, when the focus is on short and/or very
frequent $n$-grams, \textsc{Suffix}-$\sigma$ performs never
significantly worse than the other methods. It is hence robust and can
handle a wide variety of parameter choices. To substantiate this,
consider that \textsc{Suffix}-$\sigma$ could compute statistics about
arbitrary-length $n$-grams that occur at least five times (i.e.,
$\tau=5$ and $\sigma=\infty$), as reported in Figure~\ref{fig:Output},
in less than six minutes on NYT and six hours on CW.

\section{Related Work}
\label{sec:related-work}

We now discuss the connection between this work and existing
literature, which can broadly be categorized into:

\textbf{Frequent Pattern Mining} goes back to the seminal work by
Agrawal et al.~\cite{Agrawal:1993qf} on identifying frequent itemsets
in customer transactions. While the \textsc{Apriori} algorithm
described therein follows a candidate generation \& pruning approach,
Han et al.~\cite{Han:2004fk} have advocated pattern growth as an
alternative approach. To identify frequent sequences, which is a
problem closer to our work, the same kinds of approaches can be
used. Agrawal and Srikant~\cite{Agrawal:1995ve,Srikant:1996fk}
describe candidate generation \& pruning approaches; Pei et
al.~\cite{Pei:2004kx} propose a pattern-growth approach. SPADE by
Zaki~\cite{Zaki:2001ly} also generates and prunes candidates but
operates on an index structure as opposed to the original
data. Parallel methods for frequent pattern mining have been devised
both for distributed-memory~\cite{Guralnik:2004kx} and shared-memory
machines~\cite{Parthasarathy:2001vn,Zaki:2001ve}. Little work exists
that assumes MapReduce as a model of computation. Li et
al.~\cite{Li:2008fk} describe a pattern-growth approach to mine
frequent itemsets in MapReduce. Huang et al.~\cite{Huang:2010vn}
sketch an approach to maintain frequent sequences while sequences in
the database evolve. Their approach is not applicable in our setting,
since it expects input sequences to be aligned (e.g, based on time)
and only supports document frequency. For more detailed discussions,
we refer to Ceglar and Roddick~\cite{Ceglar:2006qa} for frequent
itemset mining, Mabroukeh and Ezeife~\cite{Mabroukeh:2010fk} for
frequent sequence mining, and Han et al.~\cite{Han:2007ly} for
frequent pattern mining in general.

\textbf{Natural Language Processing \& Information Retrieval.} Given
their role in NLP, multiple
efforts~\cite{Banerjee:2003uq,Ceylan:2011ly,Federico:2008fk,Huston:2011nx,Stolcke:2002kx}
have looked into $n$-gram statistics computation. While these
approaches typically consider document collections of modest size,
recently Lin et al.~\cite{Lin:2010uq} and Nguyen et
al.~\cite{Nguyen:2007uq} targeted web-scale data. Among the
aforementioned work, Huston et al.~\cite{Huston:2011nx} is closest to
ours, also focusing on less frequent $n$-grams and using a cluster of
machines. However, they only consider $n$-grams consisting of up to
eleven words and do not provide details on how their methods can be
adapted to MapReduce. Yamamoto and Church~\cite{Yamamoto:2001fk}
augment suffix arrays, so that the collection frequency of substrings
in a document collection can be determined efficiently. Bernstein and
Zobel~\cite{Bernstein:2006uq} identify long $n$-grams as a means to
spot co-derivative documents. Brants et al.~\cite{Brants:2007ys} and
Wang et al.~\cite{2010:NAACLWang} describe the $n$-gram statistics
made available by Google and Microsoft,
respectively. Zhai~\cite{Zhai:2008fk} gives details on the use of
$n$-gram statistics in language models. Michel et al.~\cite{Michel:ys}
demonstrated recently that $n$-gram time series are powerful tools to
understand the evolution of culture and language.

\textbf{MapReduce Algorithms.} Several efforts have looked into how
specific problems can be solved using MapReduce, including all-pairs
document similarity~\cite{Lin:2009fk}, processing relational
joins~\cite{Okcan:2011vn}, coverage
problems~\cite{Chierichetti:2010zr}, content
matching~\cite{Morales:2011fk}. However, no existing work has
specifically addressed computing $n$-gram statistics in MapReduce.

\section{Conclusions}
\label{sec:conclusions}

In this work, we have presented \textsc{Suffix}-$\sigma$, a novel
method to compute $n$-gram statistics using MapReduce as a platform
for distributed data processing. Our evaluation on two real-world
datasets demonstrated that \textsc{Suffix}-$\sigma$ outperforms
MapReduce adaptations of \textsc{Apriori}-based methods significantly,
in particular when long and/or less frequent $n$-grams are
considered. Otherwise, \textsc{Suffix}-$\sigma$ is robust, performing
at least on par with the best competitor. We also argued that our
method is easier to implement than its competitors, having been
designed with MapReduce in mind. Finally, we established our method's
versatility by showing that it can be extended to produce
maximal/closed $n$-grams and perform aggregations beyond occurrence
counting.



\end{document}